\documentclass[12pt,preprint]{aastex}

\begin{document}

\title{Evolution of the Solar Nebula. IX. Gradients in the Spatial 
Heterogeneity of the Short-Lived Radioisotopes $^{60}$Fe and $^{26}$Al
and the Stable Oxygen Isotopes}

\author{Alan P.~Boss}
\affil{Department of Terrestrial Magnetism, Carnegie Institution of 
Washington, 5241 Broad Branch Road, NW, Washington, DC 20015-1305}
\email{boss@dtm.ciw.edu}

\begin{abstract}

 Short-lived radioisotopes (SLRI) such as $^{60}$Fe and $^{26}$Al
were likely injected into the solar nebula in a spatially
and temporally heterogeneous manner. Marginally gravitationally
unstable (MGU) disks, of the type required to form gas giant planets,
are capable of rapid homogenization of isotopic heterogeneity
as well as of rapid radial transport of dust grains and gases
throughout a protoplanetary disk. Two different types of new models 
of a MGU disk in orbit around a solar-mass protostar are presented.
The first set has variations in the number of terms in the spherical
harmonic solution for the gravitational potential, effectively
studying the effect of varying the spatial resolution of the gravitational
torques responsible for MGU disk evolution. The second set explores the
effects of varying the initial minimum value of the Toomre $Q$
stability parameter, from values of 1.4 to 2.5, i.e., toward 
increasingly less unstable disks. The new models show that the basic
results are largely independent of both sets of variations.
MGU disk models robustly result in rapid mixing of initially highly 
heterogeneous distributions of SLRIs to levels of $\sim$ 10\% in both 
the inner ($<$ 5 AU) and outer ($>$ 10 AU) disk regions, and to
even lower levels ($\sim$ 2\%) in intermediate regions, where 
gravitational torques are most effective at mixing. These gradients
should have cosmochemical implications for the distribution of SLRIs
and stable oxygen isotopes contained in planetesimals (e.g., comets) 
formed in the giant planet region ($\sim$ 5 to $\sim$ 10 AU) 
compared to those formed elsewhere.

\end{abstract}


\section{Introduction}

 The short-lived radioisotope (SLRI) $^{60}$Fe appears to have been 
synthesized in a Type II supernova (Mostefaoui, Lugmair, \& Hoppe 2005;
Tachibana et al. 2006) and injected into the presolar cloud 
(Boss et al. 2008, 2010; Boss \& Keiser 2010) from the same massive 
star that is likely to be the source of the bulk of the solar 
nebula's $^{26}$Al (Limongi \& Chieffi 2006; Sahijpal \& Soni 2006).
Given the injection of SLRIs into the presolar cloud by 
Rayleigh-Taylor fingers (Boss et al. 2008, 2010; Boss \& Keiser 2010),
it might be expected that the SLRIs would be initially highly spatially 
and temporally heterogeneous in their distribution in the solar nebula. 
However, the nearly identical Fe and Ni isotopic compositions of iron 
meteorites, chondrites, and the Earth require that the injected $^{60}$Fe 
must have been mixed to less than 10\% heterogeneity in the solar nebula 
(Dauphas et al. 2008). A similar constraint arises from the need to 
preserve the use of $^{26}$Al as an accurate nebular chronometer 
(e.g., Thrane, Bizzarro, \& Baker 2006), while simultaneously allowing 
for the spread of stable oxygen isotope ratios (Lyons \& Young 2005; 
Lee, Bergin, \& Lyons 2008). Three-dimensional hydrodynamical models 
of the evolution of a marginally gravitationally unstable (MGU) solar 
nebula have shown that mixing of initially highly heterogeneous 
distributions of SLRIs can indeed reduce the level of heterogeneity to 
$\sim$ 10\% or lower in less than 1000 yrs (Boss 2004a, 2006, 2007, 2008).

 The discovery of refractory grains among the particles collected from
Comet 81P/Wild 2 by the Stardust spacecraft (Brownlee et al. 2006;
Simon et al. 2008; Nakamura et al. 2008)
provided the first ground truth for large-scale transport of materials 
formed in high temperature regions close to the protosun outward 
to the comet-forming regions of the solar nebula. One refractory
particle found by Stardust, Coki, has an age $\sim$ 1.7 Myr younger 
(Matzel et al. 2010) than that of calcium, aluminum-rich inclusions (CAIs),
implying that outward radial transport continued for millions of years after
CAI formation. Measurements of the iron sulfide content of particles from 
Wild 2 imply that over half of the comet's mass derived from the inner solar
nebula (Westphal et al. 2009). Similar hydrogen, nitrogen, and 
oxygen isotopic anomalies occur in both 
primitive meteorites and in cometary dust particles, implying that 
meteorites and comets both formed from the same basic mixture of disk 
material (Busemann et al. 2009; Al\'eon et al. 2009).

 Observations of disks around young stars often find evidence for 
crystalline silicate grains at distances ranging from inside 
3 AU to beyond 5 AU, in both the disk's midplane and its surface 
layers (e.g., Mer\'in et al. 2007). Only the most massive AGB stars
produce significant quantities of crystalline grains (Speck et al. 2008);
grains in the interstellar medium are primarily amorphous.
Crystalline silicate grains could have been produced through thermal 
annealing of amorphous grains (e.g., Sargent et al. 2009a,b)
by the hot disk temperatures reached only within the 
innermost disk, well inside 1 AU. Crystalline and amorphous silicate 
grains in most protoplanetary disks may thus have experienced the same 
large-scale transport phases as the refractory particles found in Wild 2.

 While accretion disk models driven by a generic turbulent viscosity 
have long been invoked as a means to explain large-scale transport 
(Gail 2001, 2002, 2004; Tscharnuter \& Gail 2007; Ciesla 2007, 2008, 2009, 
2010a,b; Birnstiel et al. 2009; Hughes \& Armitage 2010; Heinzeller et al. 
2011; Jacquet et al. 2011), the detailed physics behind $\alpha$-viscosity 
remains unclear, especially considering that the magneto-rotational
instability (MRI) often assumed to be the source of the $\alpha$-viscosity 
is unable to drive disk evolution in the magnetically dead midplane 
regions (e.g., Matsumura \& Pudritz 2006) of most interest for 
planetary formation. Objections have also been raised to the assumption
that angular momentum transport in a MRI-driven disk can be described by
the standard model for $\alpha$-viscosity (Pessah, Chan, \& Psaltis 2008).
In contrast, a MGU disk presents a self-consistent
mechanism for studying mixing and transport in protoplanetary disks, with
no free parameters beyond the initial choice of a disk massive
enough, and cold enough, to be MGU.

 We present here a new set of three-dimensional MGU disk models 
similar to those studied previously (e.g., Boss 2004a, 2006, 2007, 
2008), but with several variations intended to test the robustness of 
the conclusions about mixing and transport of the previous models.
Given that a MGU disk appears to be a likely requirement for the formation
of gas giant planets, by either core accretion (e.g., Inaba et al. 2003; 
Chambers 2008) or by disk instability (e.g., Boss 2010), these MGU models 
are intended to learn what cosmochemical consequences might also
derive from such phases of rapid disk evolution.

\section{Numerical Methods}

 The disk evolution calculations were performed with a numerical code
that uses finite differences to solve the three-dimensional equations 
of hydrodynamics, radiative transfer, and the Poisson equation for the
gravitational potential. The code is the same as that used in the
previous studies of mixing and transport in disks 
(Boss 2004a, 2006, 2007, 2008). The code has been shown to
be second-order-accurate in both space and time through convergence testing
(Boss \& Myhill 1992). The equations are solved on a spherical coordinate
grid. The number of grid points in each spatial direction is: $N_r = 51$,
$N_\theta = 23$ in $\pi/2 \ge \theta \ge 0$, and $N_\phi = 256$. This
relatively low degree of numerical spatial resolution (compared to
high resolution disk instability models, e.g., Boss 2010) was chosen
in order to evolve the disks as far forward in time as possible 
in several years of computing on a dedicated workstation. The radial 
grid is uniformly spaced between 1 and 10 AU, with boundary conditions 
at both the inner and outer edges chosen to absorb radial velocity
perturbations. The $\theta$ grid is compressed into the midplane to ensure
adequate vertical resolution ($\Delta \theta = 0.3^o$ at the midplane).
The $\phi$ grid is uniformly spaced, to prevent any bias in the azimuthal
direction. The central protostar wobbles in response to the growth of
nonaxisymmetry in the disk, thereby rigorously preserving the location of
the center of mass of the star and disk system. The number of terms in the
spherical harmonic expansion for the gravitational potential of the disk
is $N_{Ylm} = 32$ for the standard model, but has been varied to
$N_{Ylm} = 16$ or 48 in certain models, as noted below. The Jeans length
constraint (e.g., Boss 2010) is monitored throughout the evolutions 
to ensure that the spatial resolution remains adequate.

 As in Boss (2004a, 2006, 2007, 2008), the models treat radiative transfer 
in the diffusion approximation, which should be valid near the disk midplane
and throughout most of the disk, because of the high vertical optical
depth. The energy equation is solved explicitly in conservation law form, 
as are the four other hydrodynamic equations. Artificial viscosity is 
not employed. 

 In order to follow the transport of injected SLRIs, the evolution 
of a color field is calculated (e.g., Foster \& Boss 1997; Boss 2004a). 
SLRI or oxygen anomalies that reside in the gas or in particles with sizes of 
mm to cm or smaller will remain tied to the gas over timescales of $\sim$ 
1000 yrs or so, because the relative motions caused by gas drag result in 
differential migration by distances of less than 0.1 AU in 1000 yrs, 
which is negligible compared to the distances they are transported by the 
gas in that time, justifying their representation by the color field. 
The equation of motion for the color field density $\rho_c$ is exactly 
similar to the continuity equation for the mass density $\rho$, namely

$${\partial \rho_c \over \partial t} + \nabla \cdot (\rho_c {\bf v}) = 0,$$

\noindent where ${\bf v}$ is the disk gas velocity and $t$ is the time. The 
color equation is solved in the same manner as the five other equations of 
motion, using finite differences and explicit time differencing. The total
amount of color is monitored throughout the evolution to ensure the 
reliability of the color field calculation, which is conserved in the same
manner as the disk mass is conserved. While Boss (2004a) added the 
effects of an $\alpha$-viscosity to the color equation, the present 
models assume $\alpha = 0$.

\section{Initial Conditions}

 The models consist of a $1 M_\odot$ central protostar surrounded
by a protoplanetary disk with a mass of 0.047 $M_\odot$ between 1 and
10 AU, as in Boss (2008). Figure 1 depicts an azimuthal cross section 
of the initial disk structure. The underlying disk structure is the 
same as that of the disk extending from 4 to 20 AU assumed in previous 
models (Boss 2004a, 2006, 2007). Disks with similar masses appear to be 
necessary to form gas giant planets by core accretion (e.g., Inaba et 
al. 2003; Chambers 2008) or by disk instability (e.g., Boss 2010).
Protoplanetary disks are often believed to have masses in the range of 
0.01 to 0.1 $M_\odot$ (Kitamura et al. 2002), but these disk masses may 
be underestimated by factors of up to 10 (Andrews \& Williams 2007).
More recent observations of low- and intermediate-mass pre-main-sequence stars
imply disk masses in the range of 0.05-0.4 $M_\odot$ (Isella et al. 2009).
The disk all start from an axisymmetric density distribution with 
density perturbations of 1\% in the $m = 1$, 2, 3, and 4 modes.

 In the standard model 9S of Boss (2008), the disk starts with an outer 
disk temperature $T_o = 40$ K, rising to temperatures greater than
1500 K in the inner disk, as shown in Figure 2, based on the
disk midplane temperature profiles calculated by Boss (1996).
This initial temperature distribution leads to a minimum in the 
initial Toomre $Q$ value of 1.4 at the outer boundary of the
active disk at 10 AU (Figure 3). Inside $\sim$ 5 AU, $Q$ rises to 
values $>$ 10 because of the much higher disk temperatures closer 
to the protosun. A $Q$ value of $\sim$ 1.4 implies marginal instability 
to the growth of gravitationally-driven perturbations, while $Q > 10$ 
implies a high degree of stability. Five other models (Table 1) explore
the effects of different initial values of the outer disk temperature,
$T_o =$ 60, 70, 80, 100, and 120 K, resulting in initial minimum
Toomre $Q$ values of 1.8, 1.9, 2.0, 2.3, and 2.5, respectively.

 In all of these models, a color field representing SLRI is sprayed 
onto the outer surface of the disk at a radial distance centered 
on 8.5 AU (Figure 1) into a 90 degree (in the azimuthal direction) 
sector of a ring of width 1 AU, simulating the arrival of a 
Rayleigh-Taylor finger carrying SLRI (e.g., Boss et al. 2008, 2010; 
Boss \& Keiser 2010). While Boss (2004a) found that models with
injection into a 90 degree sector evolved in much the same way
as models with injection into a 360 degree ring, it would be interesting
in the future to investigate injection into sectors smaller than 90 
degrees, which might better represent the extent of Rayleigh-Taylor
fingers upon impact with the solar nebula.

\section{Results}

 We present here the results of eight new models (Table 1). 
Model 1.4 is the continuation in time of model 9S from Boss (2008).
Models 16 and 48 are identical to the model 1.4, except for having 
the number of terms in the spherical harmonic expansion for the 
gravitational potential solver changed from $N_{Ylm} = 32$ to either 16 or 48. 
Since MGU disks evolve solely as a result of gravitational torques, these 
changes have the effect of varying the numerical resolution.
Models 1.8, 1.9, 2.0, 2.3, and 2.5 are identical to model 1.4, except 
for starting with a more gravitationally stable disk, as quantified 
by initial minimum Toomre Q values of 1.8, 1.9, 2.0, 2.3, and 2.5,
respectively, instead of 1.4. 

\subsection{Evolution of the Disk}

 Model 2.5 serves to demonstrate the common evolutionary outcome of
all of these MGU disks, and so will be described in some detail.

 Figures 4 and 5 display the midplane density and temperature
distributions for model 2.5 after 1548 yr of evolution. While
model 2.5 is initially the most gravitationally stable model
calculated to date, it is clear that even in model 2.5, the growth
of spiral arms in the outer disk drives the formation of non-axisymetric
structures even in the highly gravitationally stable inner disk.
Figure 4 shows that the outer spiral arms have driven a one-armed
trailing spiral (the disks rotate in the counter-clockwise direction)
right down to the central protostar. Such an inner one-armed spiral
is a likely means for accomplishing the thermal processing in a
shock front that appears to be necessary to account for chondrule
formation (Boss \& Durisen 2005). Figure 5 shows that while the
hot inner disk remains thermally axisymmetric, non-axisymmetric
temperatures accompany the spiral features in the cooler outer disk.
The temperature distribution in the outer disk is nevertheless more 
uniform than the density distribution, because of the restriction that 
the temperature cannot drop below the initial value of 120 K for model 
2.5, which is considerably higher than that ($\sim$ 50 K) to be expected 
for the solar nebula (Boss 1996). Restrictions on the temperature
fields are imposed in order to err on the conservative side with
respect to the possible growth of clumps in disk instability models 
of gas giant planet formation (e.g., Boss 2010).

\subsection{Evolution of the Color Field}

 Figure 6 shows the midplane distribution of the color field for
model 2.5 after only 112 yr of evolution, or 3.6 orbital
periods at the 10 AU edge of the disk. By this early time,
the color has already been transported downward to the midplane
from the disk's surface (Figure 1) as well as azimuthally around 
the disk, to all azimuthal angles, compared to the initial
90 degree sector. Given the initial injection at 8.5 AU, where 
the orbital period is 25 yr, onto the disk's surface in a sector 
between 9 o'clock and 6 o'clock, by 112 yr any material orbiting 
at 8.5 AU should have made 4.5 orbits, so in the absence of any 
other effects, the color field should have a maximum between 3 o'clock
and 12 o'clock in Figure 6 (the disk rotates in the counterclockwise
direction). The fact that the color field's maximum is instead 
at 12 o'clock is due to color field maximum having been transported 
inward by 112 yr to a distance of $\sim$ 8.0 AU, where the orbital
period is only 22 yr, allowing the color field maximum to have
orbited further than the prediction based on an 8.5 AU orbit.

 Figure 6 also shows that some of the color field has
already been transported to the inner edge of the disk, where
is it beginning to accrete onto the central protostar.
The rapid vertical transport appears to be due at least in
part to convective-like motions driven by the 
superadiabatic vertical temperature gradients between the disk's 
midplane and its upper layers that develop as the disk evolves
(Boss 2004b). In addition, the global mass transport driven
in the midplane by the spiral arms (Figure 4) clearly results
in rapid radial and azimuthal transport. Figure 7 shows the
same model 2.5 after another 1436 yr of evolution (same time
as Figure 4), where it can be seen that the color field has
been preferentially trapped in the clumps and spiral arms
in the disk midplane, where the gas density is highest and
where the gravitational potential minima occur. Significant
amounts of color have been accreted by the protostar, and have also 
piled up at the outer boundary of the active computational
disk, indicative of the fact that MGU disks drive transport both
inwards and outwards in the disk. By the time that model 2.5
was stopped, after 1548 yr, 25\% of the initial color field
had been accreted by the central protostar. For comparison,
model 1.8 by that same time had accreted 42\% of the color
field onto the protostar, indicative of faster inward accretion
rates in a lower $Q$ disk, as expected.

 A comparison of Figures 6 and 7 would seem to indicate, however,
that the color field is becoming more heterogeneously distributed
with time, rather than becoming homogenized. This is misleading, 
because what is important for cosmochemical heterogeneity is
the relative abundances of SLRIs (e.g., the abundance ratio
$^{26}$Al/$^{27}$Al), not their absolute abundance (e.g., the
number of $^{26}$Al atoms). Figures 8 and 9 display the color field
density divided by the gas field density, i.e., $^{26}$Al/$^{27}$Al
(most of the $^{27}$Al resides in the pre-injection cloud), for
model 2.5 at the same two times as shown in Figures 6 and 7.
Figure 8 shows that 112 yr after injection, the color field is
strongly heterogeneous, having been freshly injected, whereas
by 1548 yr (Figure 9), the MGU disk has rather thoroughly
homogenized the color field with respect to the underlying
gas density field. The absence of contours in most of Figure 9
shows that the color field has been homogenized to less than
25\% variations for the entire disk, with the exception of
the innermost few AU.
 
\subsection{Evolution of the Color Field Dispersion}

 In order to follow the extent to which the color field is
homogenized in MGU disks, the time evolution of the dispersion
of the color field is computed. The dispersion is defined to
be the square root of the sum of the squares of the color field 
divided by the gas density, subtracted from the mean value of the 
color field divided by the gas density, where the sum is taken over 
the midplane grid points and is normalized by the number of grid points 
being summed over (e.g., Boss 2006, 2007, 2008). The sum excludes
the regions closest to the inner and outer disk boundaries, in order
to minimize the effect of the artificial disk boundaries.

 Figure 10 shows the time evolution of the dispersion for models
1.4, 16, and 48, where the only difference in the models is in
the number of terms retained in the spherical harmonic expansion
for the gravitational potential solver, 32, 16, and 48, respectively.
All of the models start off from a highly heterogeneous initial
condition, with a dispersion much greater than 1. Figure 10 shows 
that the dispersion in all three models thereafter behaves in
much the same way, though the rate at which the disk is homogenized
depends slightly on the gravitational potential resolution.
In model 48, the dispersion decreases somewhat faster than in
model 1.4, which drops somewhat faster than in model 16, as expected,
as the homogenization is being performed by the spiral arms, which
become increasingly vigorous as the spatial resolution is
effectively improved. However, Figure 10 shows that, at least
over the range investigated here, the degree of spatial resolution
of the spiral arms does not have a significant effect on the
evolution of the dispersion, i.e., these MGU disk models appear
to be adequately resolved.

 Figure 11 presents the evolution of the dispersion for models
1.4, 1.8, and 2.5. Again the behaviors are quite similar to
each other, though model 2.5 clearly is lagging behind models
1.4 and 1.8 in achieving homogenization. Even still, the dispersion
in model 2.5 drops down to a low value within $\sim$ 500 yr of 
evolution, compared to within $\sim$ 300 yr for models 1.4 and 1.8.
Again, the rate at which a MGU disk homogenizes an initial
heterogeneity appears to be relatively independent of the
initial degree of gravitational instability, at least for the
range of disk models examined here. Furthermore, the degree
to which the dispersion is reduced also appears to be relatively
independent of the initial Toomre $Q$ value.

 Figures 10 and 11 imply that the dispersion drops rapidly
to equally low values for MGU disks, but that the dispersion does not
fall to zero. In order to investigate the extent to which
the dispersion might reach an asymptotic or steady level,
several of the models were evolved for considerably longer
periods of time than in the previous 10 AU disk models (Boss 2008),
none of which were calculated for more than 1500 yr.

 Figure 12 shows the evolution of the dispersion for model 1.4
for nearly 3000 yr, with the dispersion being plotted separately
for both the inner (1 to 5.5 AU) and outer (5.5 to 10 AU) disks.
Figure 12 shows that over time periods of 3000 yr, the dispersion
does not disappear, but remains finite at a level of less than
$\sim$ 15\%. Figure 13 replots the same data as Figure 12, but
with the vertical scale altered to reveal the finer details
of the evolution of the dispersion in model 1.4. Clearly 
mixing and transport in a MGU disk are chaotic processes, with
transient periods of variable heterogeneity, though to some
extent this apparent variability is caused by the transient
behavior of the underlying gas density field, as it is being
accreted by the central protostar, or is being piled up artificially
at the outer disk boundary. Since the gas density field is a
normalizing factor in the calculation of the dispersion, the
chaotic behavior seen in Figure 13 cannot be attributed solely
to mixing and transport of the color field. Nevertheless,
the gross features evident in Figure 13 are clearly driven by 
the color field evolution, as shown by previous models, where,
e.g., very low gas density regions were excluded from the
dispersion calculation (Boss 2007).
 
 Figure 13 shows that for model 1.4, the dispersion is typically
significantly lower in the outer disk than in the inner disk.
Figures 14, 15, and 16 display the evolution of the dispersions
for models 16, 1.8, and 2.5, respectively, showing that this same
trend of lower dispersion in the outer disk compared to the
inner disk is independent of both the disk's effective gravitational 
spatial resolution (model 16) and the disk's initial degree
of gravitational instability (models 1.8 and 2.5). Note that
while Figures 14 and 15 imply dispersions that are increasing
somewhat after 1500 yr, a comparison with Figure 12 for model 1.4
shows that such increases are likely only transitory, and that if
calculated further, the dispersions for models 16 and 1.8 would
also behave like that of model 1.4.

 It is useful to compare these results with those from model 15S
of Boss (2006), which is identical to the present model 1.4 (9S) 
except for having the injection occur at 15 AU on the surface of a
disk extending from 4 to 20 AU. Model 15S thus explores the
evolution of the dispersion at greater radial distances than 
the present models do. Figure 17 shows the evolution
of the dispersion for model 15S over a time period up to
6000 yr, showing that in this disk, the dispersion hovers
around values of $\sim$ 10\% to 15\% in the outer disk (12 to 20 AU)
and around values of $\sim$ 1\% to 2\% from 4 to 12 AU. In a
sense, this is in the reverse sense of the models in this paper,
but in reality this is entirely consistent: MGU disks result in
the most efficient homogenization in the region where the gas
giant planets are trying to form, roughly 5 to 10 AU, where the
gravitational torques are the strongest, while the degree of
homogenization attained is measurably lesser in the regions both interior
to (i.e., inside $\sim$ 5 AU) and exterior to (i.e., outside $\sim$ 10 AU)
the gas giant planet formation region.

 It is also interesting to determine the extent to which the
MGU disk evolves during the time periods covered by these models. 
Figure 18 shows the time evolution of the mass accretion rate from 
the disk onto the central protostar for model 1.8, which effectively
means any disk mass that passes through the inner disk boundary
at $\sim$ 1 AU. Figure 18 shows that mass accretion rate in these
MGU disks is highly variable, with mass accretion rates that
can vary by almost 10 orders of magnitude over time periods of
a few hundred years. However, the initially highly variable
accretion rates must be in part a result of starting off the
evolutions from a nearly axisymmetric, yet MGU disk: the disk
must then quickly adust to the fact that it should be highly
non-axisymmetric. [Note that Boss (2007) studied injection onto
disks that had already been evolving for several hundred years, and
found similar evolutions of the color field dispersion as in the
present models, so the choice of the initial disk model does not
seem to have a significant effect on the mixing and transport
processes.] After several hundred years of evolution, the mass
accretion rate becomes less chaotic, though it still varies 
by factors of $10^2$ to $10^4$ over short time scales. The
evolution of the dispersion during this latter time period
appears to be more or less independent of this decrease in
the mean mass accretion rate (Figures 12 and 13), implying
that the degree of homogeneity achieved is not strongly dependent
on the central protostar's mass accretion rate.

 During the model 1.8 evolution for 2244 yr, the mean mass
accretion rate drops from about $\sim 10^{-5} M_\odot$ yr$^{-1}$
to $\sim 10^{-6} M_\odot$ yr$^{-1}$. Even the latter rate
is quite high, however, and ultimately unsustainable for a
MGU disk. Figure 19 plots the growth in mass of the central 
protostar for model 1.8, showing that the protostar gains
$\sim 0.03 M_\odot$ during the evolution, i.e., about
60\% of the total available initial disk mass of $0.047 M_\odot$.
During the same time period, the protostar accretes 48\%
of the initial color field, a somewhat smaller fraction, given
the need for the color field to be first transported inward 7.5 AU
before it can pass inside 1 AU. Because the inner disk is
initially color free, yet begins accreting onto the protostar
as soon as the evolution starts, the accreted material is
initially color-free, which could reduce the RSS dispersion even 
if no mixing were occurring. By 268 yr for model 1.8, 20\% of the 
disk mass has entered the protostar, while only 1\% of the 
color field has been accreted. As a result, a portion of the sharp 
initial declines in the dispersions seen in Figure 11 could be due 
to the initial accretion of necessarily color-free disk gas, 
given the color field injection at 8.5 AU. However, this initial
effect must be a small portion of the total drop seen in Figure 11,
since the dispersion is defined to be the root of the sum of the 
squares [RSS] of the differences from the mean of the color field 
density divided by the gas density, and in regions where the color
field is zero, the gas density is irrelevant, as their ratio is zero
regardless of the value of the gas density. Hence color-free regions
enter into the dispersion with the same weight, regardless of their gas 
density. This conclusion is supported by the fact that the same sharp 
drops in dispersion are seen in models identical to model 1.4 except
for having the color field injected at 2 AU rather than at 8.5 AU
(Boss 2008), so that the color field is much more quickly accreted
onto the protostar. Hence this initial color-free accretion phase
does not appear to be the dominant effect seen in Figure 11; rather,
mixing in the MGU disk dominates.

 The MGU phase for model 1.8 is winding down toward the 
end of the evolution, as can be seen in Figure 19 by the approach
of the central protostar mass toward an asymptotic value.
MGU disks thus appear to evolve in an inherently transient
manner: some process, such as MRI instability in the ionized
surface or outer disk layers, or onging accretion from the
placental dense molecular cloud core, may lead to a pile-up of disk
mass that initiates a new phase of MGU evolution.

\section{Discussion}

\subsection{Cosmochemical Constraints}

 Cosmochemical studies of meteoritical and terrestrial samples have
constrained the degree of isotopic heterogeneity in the inner solar
system, and hence of the inner solar nebula at the time of formation
of the first solids. The SLRI $^{26}$Al appears to have been
homogeneously distributed to within percentages ranging from
5\% (Bouvier \& Wadhwa 2010) to 10\% (Villeneuve et al. 2009)
to 30\% (Schiller et al. 2010). Similarly, the SLRI $^{60}$Fe appears to 
have been homogeneously distributed to within percentages ranging from
10\% (Dauphas et al. 2008) to 15\% (Moynier et al. 2009).
The SLRI $^{41}$Ca appears to have been even much more
homogeneously distributed to within 0.1\% (Simon \& DePaolo 2010).
Similarly, $^{54}$Cr appears to have been homogeneously distributed 
to within 0.1\% (Dauphas et al. 2010; Yamakawa et al. 2010;
Qin et al. 2010; Yamashita et al. 2010), as was Ti (Leya et al. 2009;
Trinquier et al. 2009). 

 Stable oxygen isotope variations, on
the other hand, vary by as much as 10\% (Clayton 1993) to 20\%
(Sakamoto et al. 2007) across the solar system. 
Variations in the stable oxygen isotope ratios measured in the rim 
of a single CAI imply that this CAI had been transported to regions 
of the solar nebula with oxygen isotope ratios varying by as much as 
$\sim$ 3\% above the most $^{16}$O-poor regions (Simon et al. 2011).

 Some of the finite heterogeneity inferred for the meteoritical
samples may have been caused by parent body processing rather than
by nebular heterogeneity (Yokoyama, Alexander, \& Walker 2011), 
implying that the nebula may have been even more homogeneous than 
these studies suggest. However, the planetary accretion process
itself tends to homogenize bulk compositions (e.g., Simon, DePaolo,
\& Moynier 2009), which works in the opposite direction of allowing
greater nebular isotopic heterogeneity. While the cosmochemical
record thus cannot be considered to be perfectly clear, the need
for a physical mechanism capable of homogenizing initial isotopic
heterogeneities to dispersions ranging from as small as 0.1\%
to as large as 10\% seems to be required. This finite level of 
residual nebular spatial heterogeneity appears to be 
related to the relatively coarse mixing achieved by spiral arms, with 
radial widths of order 1 AU, over time scales of up to $\sim$ 3000 yr,
as shown by the present and previous models (e.g., Boss 2006, 2007).

\subsection{Observational Constraints}

 Observations of the turbulent linewidths of molecular species in
protoplanetary disks imply a turbulent viscosity parameter 
$\alpha$ with a value of $\sim 0.01$ (Hughes et al. 2011). 
While Ciesla (2007) showed results for $\alpha =$ 0.002 and 0.0002,
Ciesla (2010b) studied particle trajectories in disks with 
$\alpha$ ranging from 0.01 to 0.0001, finding that in the
former disks, particles were transported rapidly through
the disk, while in the latter, transport was much slower.
Boss (2004a) estimated that the effective $\alpha$-viscosity
of the MGU disks he studied was $\sim$ 0.001. MGU disks evolve
to a certain degree in a manner similar to generic turbulent
viscosity accretion disks (Lodato \& Rice 2004, 2005), so
one could just as well imagine the driving mechanism for
generic accretion disk evolution to be due to MGU as to MRI.

 Crystalline mass fractions in 1 to 2 Myr-old disks do not appear to correlate
with properties such as stellar mass or luminosity, stellar accretion
rate, disk mass, or disk to star ratio, as might be expected for
any mechanism based on large-scale transport (Watson et al. 2009).
Other arguments have been presented as well in favor of local 
heating processes rather than large-scale transport, namely differences
in the abundances of different crystalline silicate species between
the hot innermost disk ($<$ 1 AU) and the cooler outer disk beyond 
5 AU (Bouwman et al. 2008). However, the present models show
that the extent of mixing and transport in MGU disks is not
strongly dependent on the initial Toomre $Q$ stability parameter,
implying that any variation with observational parameters such
as stellar mass, disk mass, etc., may also not be as strong
as might otherwise be expected.

 Miller et al. (2011) detected an FU Orionis outburst in a classical
T Tauri star, a Class II-type object where the disk is typically 
thought to be no more massive that $\sim 0.01 M_\odot$. However,
this occurence of an FU Orionis outburst in LkH$\alpha$ 188-G4 shows
that even the disks around Class II-type objects can be subject
to transient instabilities that produce high mass acretion rates
onto the central protostar. Gravitational instabilities of the
type studied here are the leading candidate for producing FU Orionis
outbursts (e.g., Zhu et al. 2009); the mass accretion rates shown 
in Figure 18 strongly support this candidacy.

\section{Conclusions}

 All eight of these new models evolve in much the same manner 
as model 9S of Boss (2008): the initially high degree of SLRI 
heterogeneity is lowered by mixing within $\sim$ 500 yrs to a 
dispersion of $\sim$ 10\% inside 5 AU and $\sim$ 2\% from 5 AU 
to 10 AU. Combined with previous results for disks extending to 
20 AU (Boss 2007), yielding dispersions of $\sim$ 10\% beyond 10 
AU, gradients in isotopic heterogeneity are to be expected in MGU 
disks. MGU disk models thus make a clear prediction: solids that
accreted in the gas giant planet region from $\sim$ 5 to 10 AU 
should be significantly more isotopically homogenized than those
that formed interior to, or exterior to, this region. This bold
prediction must be taken with a grain of salt, however, as it 
presumes locally closed system behavior, and does not take into account
other factors, such as the arrival of multiple Rayleigh-Taylor fingers
with varying time intervals, or ongoing photochemistry due
to self-shielding of CO molecules at the disk's surfaces.

 Given that all meteoritical samples are believed to originate
from planetestimals formed in the inner solar nebula, there may not
be any measurable cosmochemical consequence of this predicted
gradient in isotopic heterogeneity. Nevertheless, 
the robustness of the present MGU disk models supports 
the idea of the existence of a low degree of spatial heterogeneity 
of SLRIs in the inner solar nebula, consistent with the long-held belief
that cosmochemical variations in inferred initial $^{26}$Al/$^{27}$Al 
ratios can be used as accurate chronometers for the timing of planetesimal 
and planet formation events in the inner solar system. 
While the FUN inclusions show no evidence for live $^{26}$Al and so 
are thought to have formed prior to the arrival of the first 
Rayleigh-Taylor finger (Sahijpal \& Goswami 1998), most refractory 
inclusions have $^{26}$Al/$^{27}$Al ratios that scatter by no more 
than $\sim$ 10\% around the mean value of $\sim 4.5 \times 10^{-5}$ 
(MacPherson et al. 1995; Young et al. 2005). 
This level of scatter in $^{26}$Al/$^{27}$Al 
ratios appears to be consistent with the 10\% dispersion expected
in the inner regions of disks that have experienced a MGU phase.

 However, there may well be a cosmochemically observable signature of
this predicted isotopic heterogeneity gradient in comets, and in
the interplanetary dust particles (IDPs) that originate in comets.
Comets that formed in the region of the gas giant planets are believed
to have been kicked outward by close encounters with the giant
planets, either to hyperbolic orbits, or to highly eccentric orbits,
i.e., to the Oort Cloud (e.g., Paulech et al. 2010). Long-period comets 
are believed to be derived from the Oort Cloud and thus to have 
originated from the most isotopically homogeneous region of a MGU disk.
Short-period comets, on the other hand, are thought to be derived from 
the Kuiper Belt, i.e., from an origin in the outer disk (e.g., Levison 
et al. 2008), where the same high degree of isotopic homogeneity would 
not have occurred in a MGU disk. Assuming that these ideas about the 
origins of long-period and short-period comets are correct, MGU disk models 
make a clear prediction: long-period comets should be more isotopically
homogeneous than short-period comets. In addition, both types of comet 
should contain refractory particles similar to CAIs, such as the
Coki particle in Wild 2.

\acknowledgments

 I thank the referee for a number of perceptive suggestions and Sandy 
Keiser for cluster and workstation management. This research was supported 
in part by the NASA Planetary Geology and Geophysics Program (NNX07AP46G)
and by the NASA Origins of Solar Systems Program (NNX09AF62G), and 
is contributed in part to the NASA Astrobiology Institute 
(NNA09DA81A). Calculations were performed on the Carnegie Alpha 
Cluster, which was supported in part by the NSF MRI Program (AST-9976645).

\clearpage

\begin{deluxetable}{lcccc}
\tablecaption{Initial conditions and final times ($t_f$ in yr) for the models.
\label{tbl-1}}
\tablewidth{0pt}
\tablehead{\colhead{model}
& \colhead{$N_{Ylm}$}
& \colhead{$Q_{min}$}
& \colhead{$T_o$}
& \colhead{$t_f$}}
\startdata
      
16  & 16 & 1.4 &  40 & 1621 \cr

48  & 48 & 1.4 &  40 &  455 \cr

1.4 & 32 & 1.4 &  40 & 2834 \cr

1.8 & 32 & 1.8 &  60 & 2244 \cr

1.9 & 32 & 1.9 &  70 & 2258 \cr

2.0 & 32 & 2.0 &  80 &  275 \cr

2.3 & 32 & 2.3 & 100 & 1588 \cr

2.5 & 32 & 2.5 & 120 & 1548 \cr

\enddata
\end{deluxetable}

\clearpage

\begin{figure}
\vspace{-2.0in}
\plotone{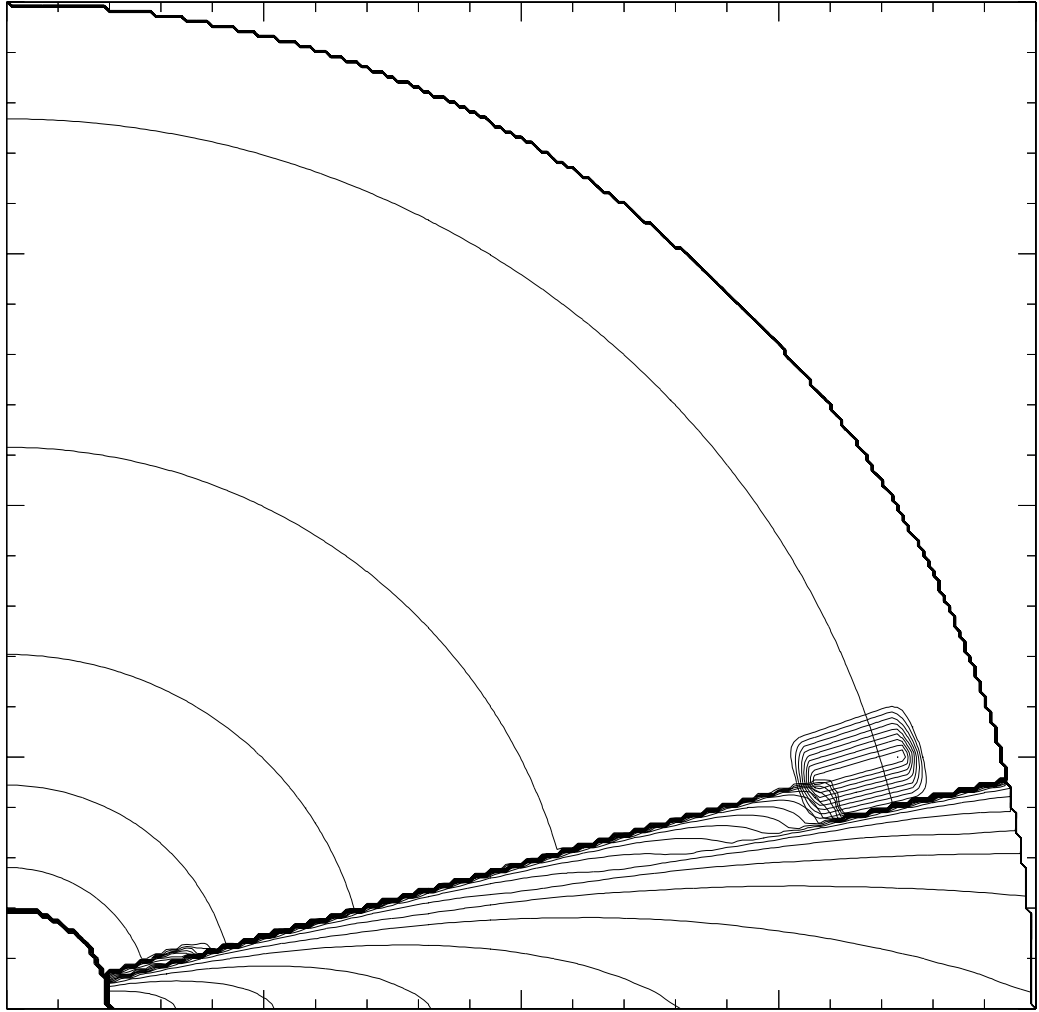}
\caption{Azimuthal disk cross-section for the initial models, showing
logarithmic contours of the gas density, with each contour representing 
a factor of 2 change in density. The numerical grid extends from
the inner boundary at a radius of 1 AU to the spherical outer boundary
at 10 AU. A solar-mass protostar lies at the lower left hand corner
of the plot. The models have reflection symmetry through the disk
midplane, which is at the bottom of the plot. The initial color
field is also displayed, represented by the linear contours just
above the disk's surface between 8 and 9 AU in radius. The color 
contours are spaced by units of 0.1 below the maximum value of 1.0.}
\end{figure}

\clearpage

\begin{figure}
\vspace{-2.0in}
\plotone{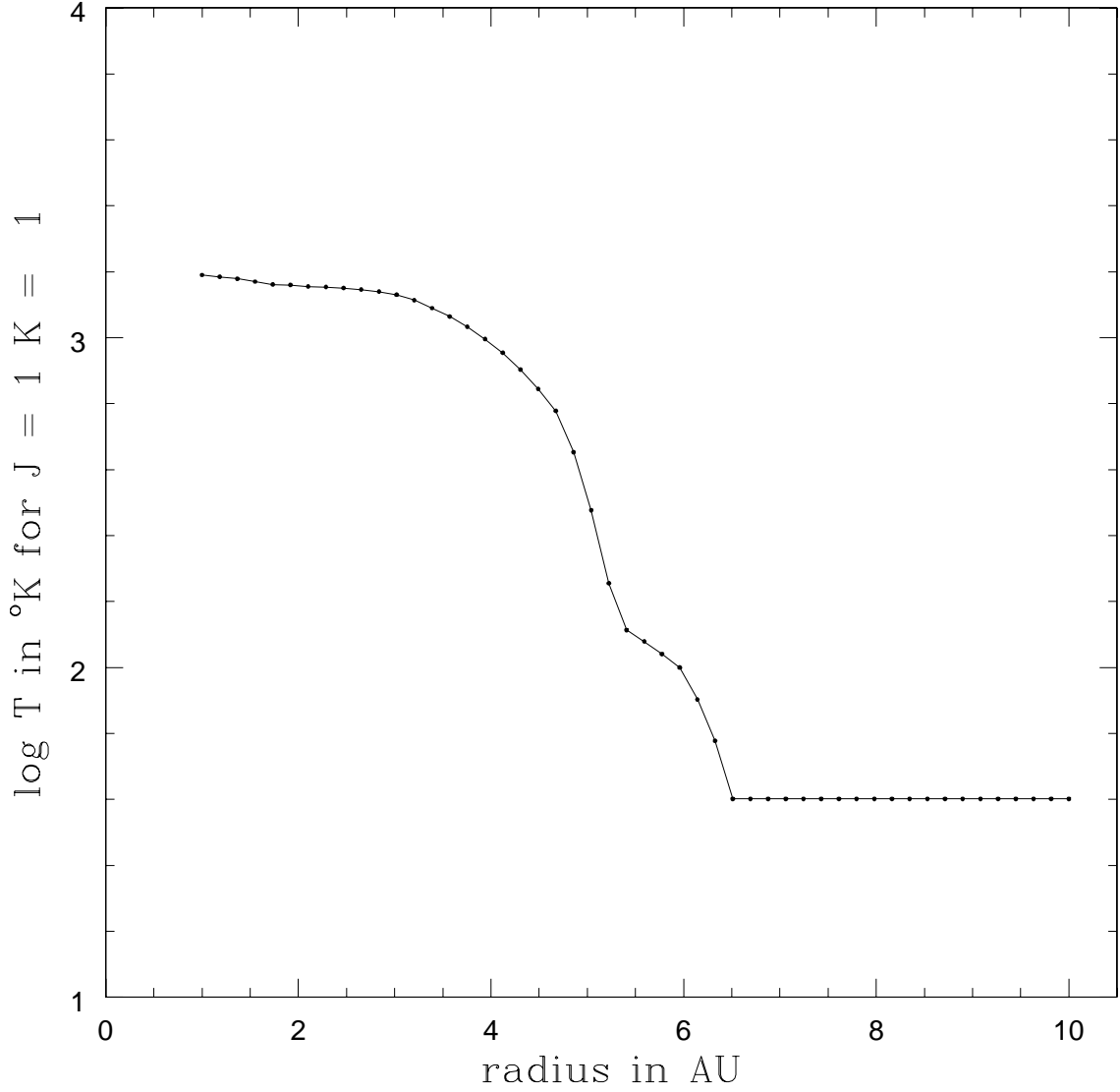}
\caption{Initial midplane temperature profile for the standard
model, model 1.4, with an outer disk temperature of 40 K. The inner
disks are too hot to be gravitationally unstable.}
\end{figure}

\clearpage

\begin{figure}
\vspace{-2.0in}
\plotone{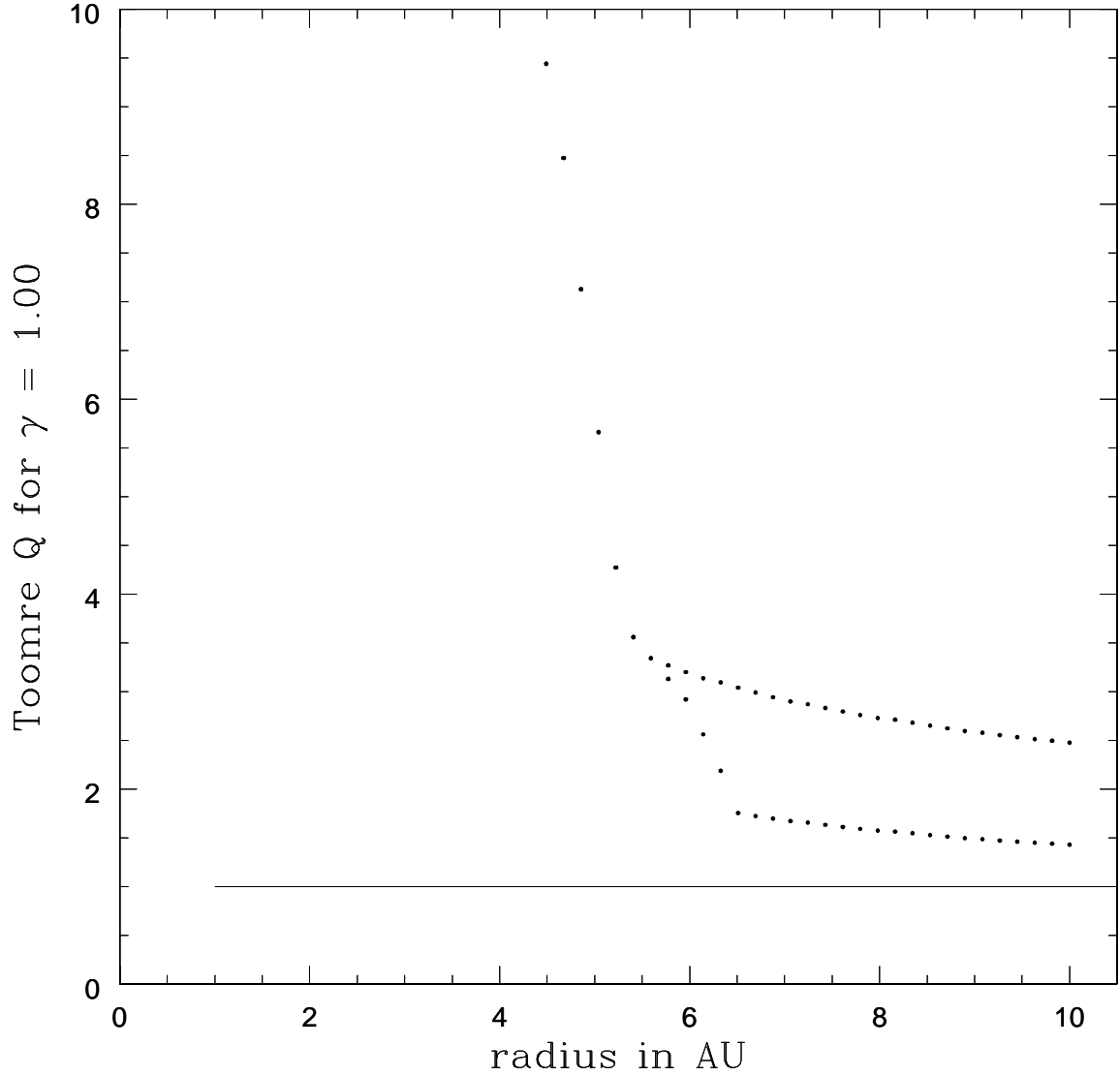}
\caption{Initial Toomre $Q$ profile for model 1.4 (lower set of dots)
and for model 2.5 (upper set of dots), compared to $Q = 1$ (solid line),
denoting a strongly gravitationally unstable disk. Model 2.5 has an
outer disk temperature of 120 K, compared to 40 K for model 1.4, 
making model 2.5 considerably less gravitationally unstable initially
than model 1.4.}
\end{figure}

\clearpage

\begin{figure}
\vspace{-1.0in}
\plotone{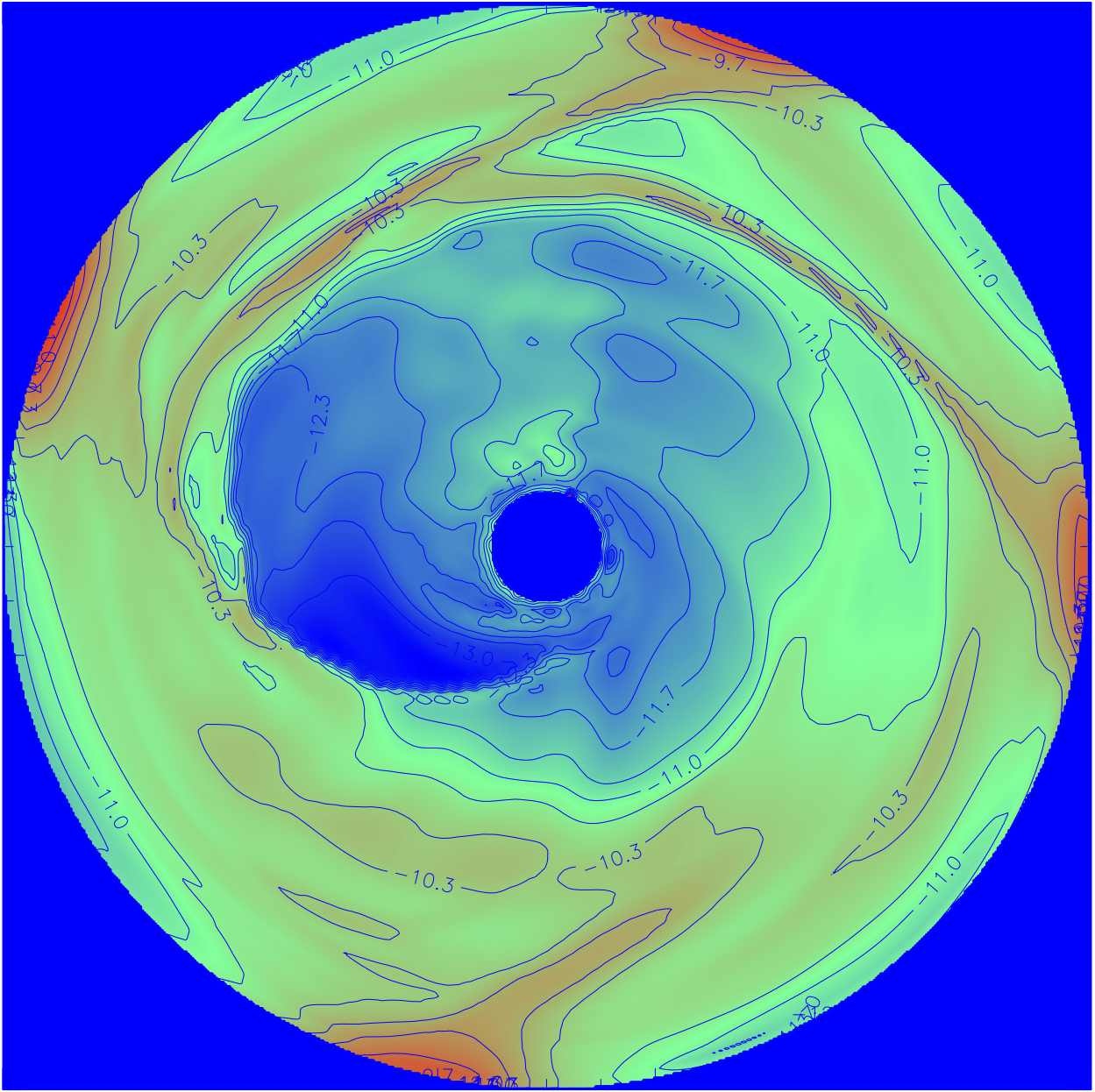}
\caption{Midplane density contours for model 2.5 after 1548 yr of evolution.
Colors span a rainbow from blue (low density) to red (high density).
Contours are spaced by factors of $\sim$ 2 in density. Region shown
is 20 AU wide; the inner boundary with a radius of 1 AU is evident.}
\end{figure}

\clearpage

\begin{figure}
\vspace{-1.0in}
\plotone{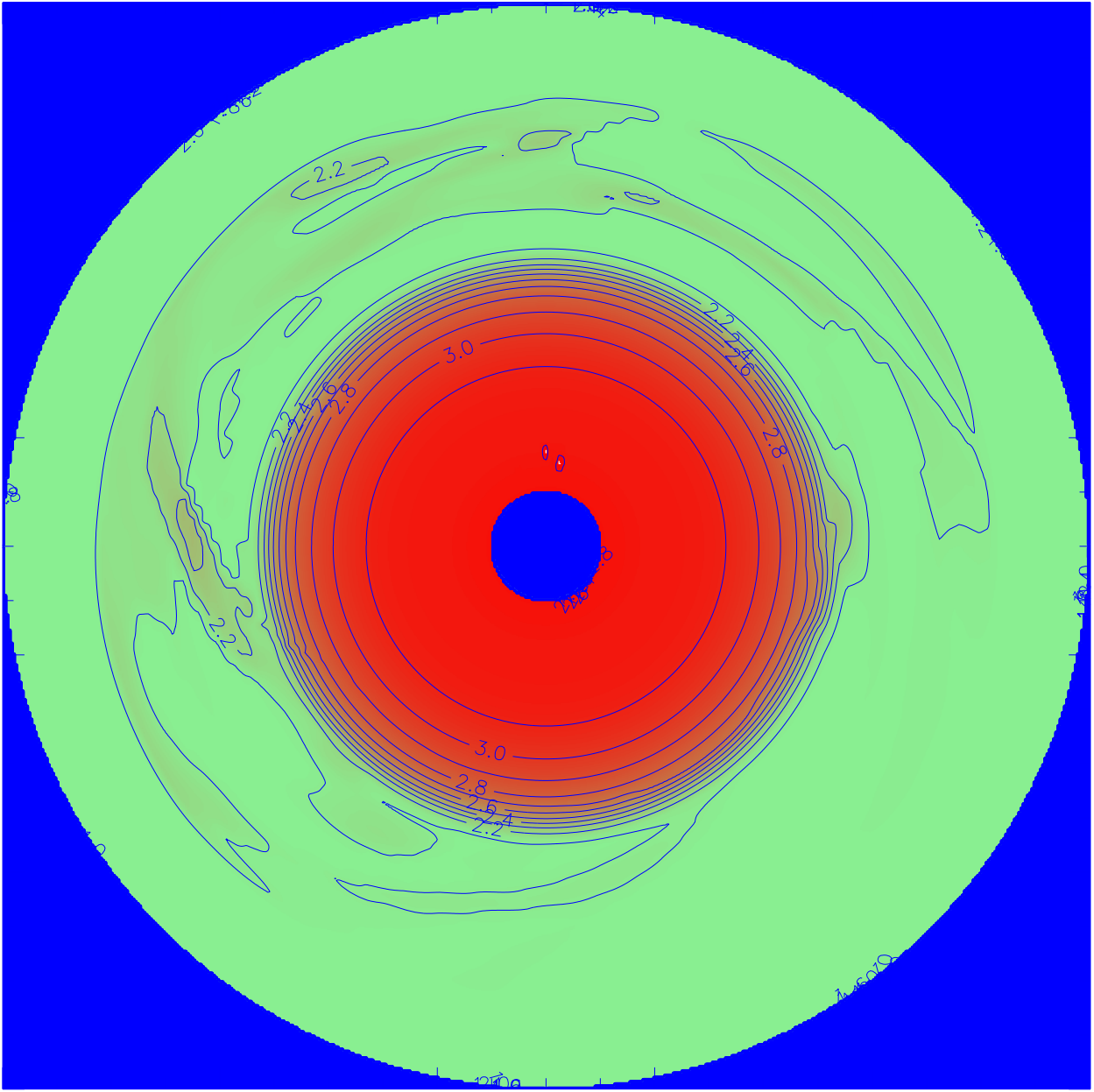}
\caption{Same as Figure 4, except showing the midplane temperature 
contours for model 2.5 after 1548 yr.}
\end{figure}

\clearpage
       
\begin{figure}
\vspace{-2.0in}
\plotone{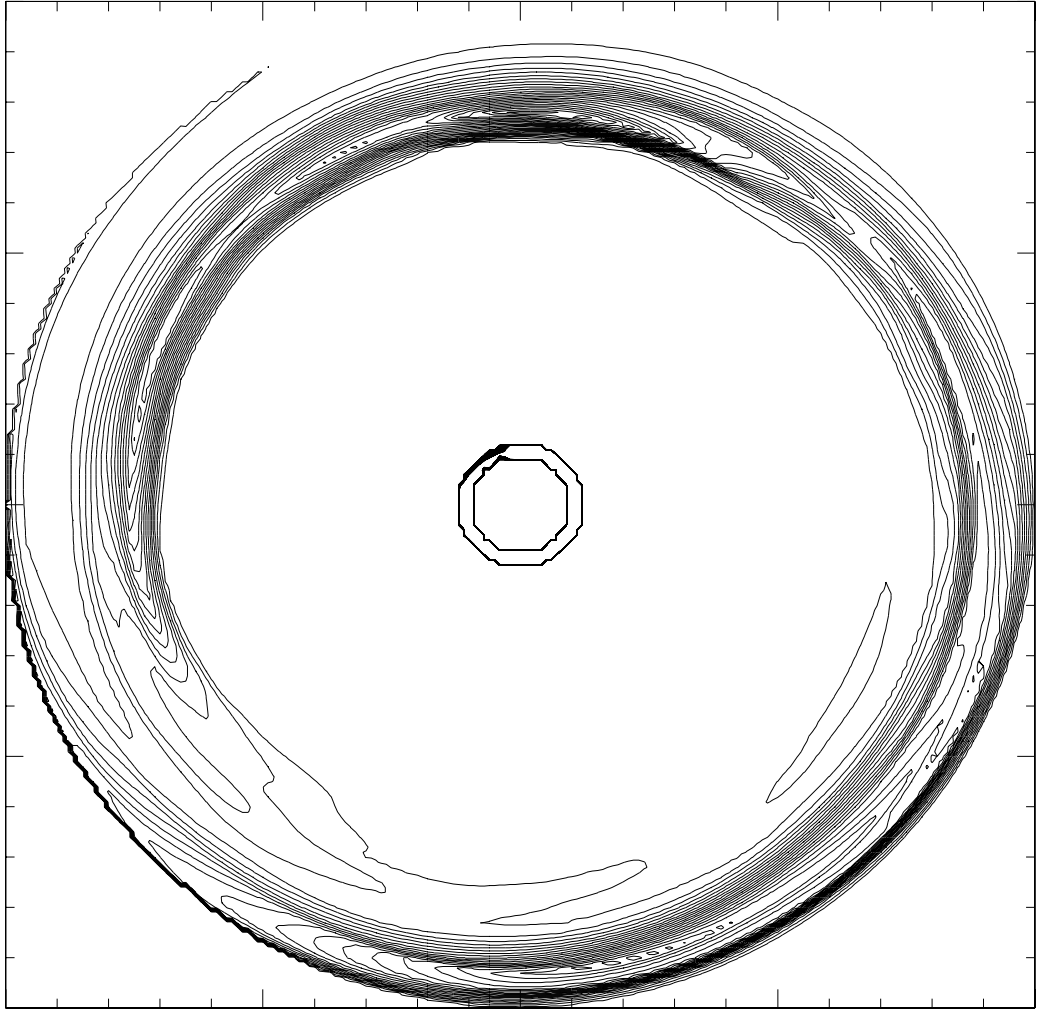}
\caption{Model 2.5 after 112 yr, showing linear contours of 
the color field density (e.g., number of atoms of $^{26}$Al cm$^{-3}$) 
in the disk midplane. Region shown is 10 AU in radius with a 1 AU radius 
inner boundary. In this Figure, the contours represent changes in the color 
field density by 0.01 units (CONDIF) on a scale normalized by the 
initial color field density of 1.0, up to a maximum value of 1.0 (CMAX). 
By this time, the color field has reached the midplane and has spread 
radially inward and outward. The maximum in the midplane color field 
of 0.25 occurs at 12 o'clock.}
\end{figure}

\clearpage

\begin{figure}
\vspace{-2.0in}
\plotone{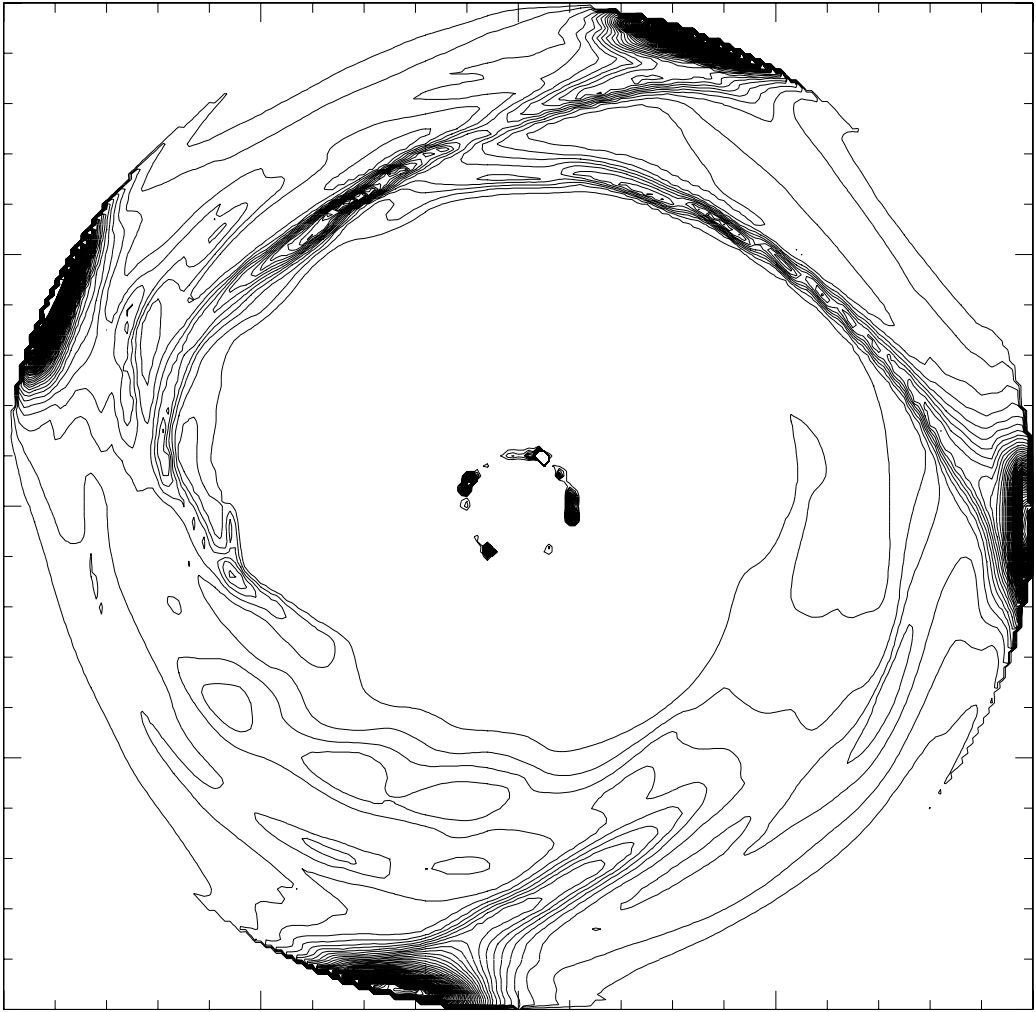}
\caption{Same as Figure 6, except after 1548 yr of evolution of
model 2.5.}
\end{figure}

\clearpage

\begin{figure}
\vspace{-2.0in}
\plotone{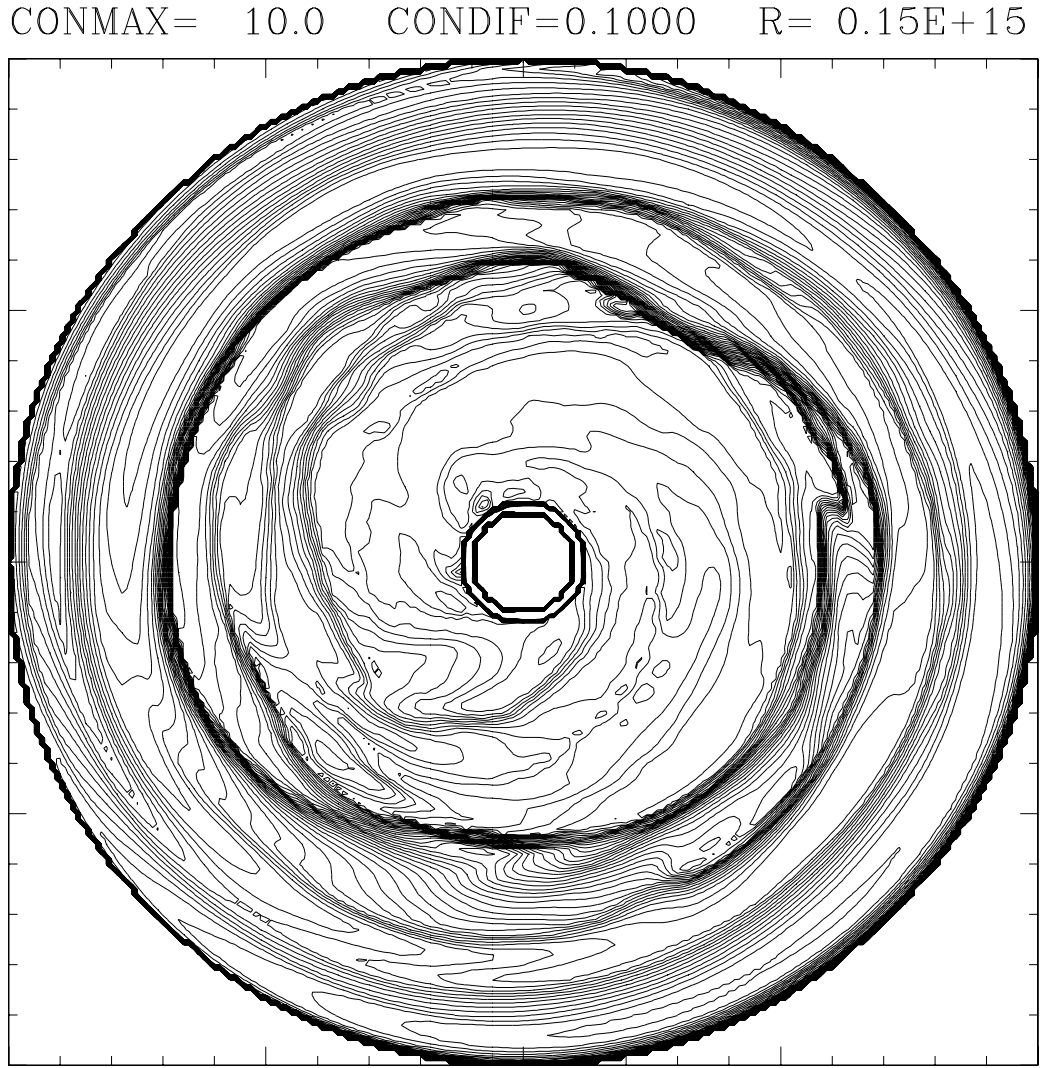}
\caption{Logarithmic contours of the color field density divided by 
the disk gas density (i.e., log of the abundance ratio $^{26}$Al/$^{27}$Al) 
for model 2.5 at a time of 112 yr. Contours represent changes by 
factors of 1.26 up to a maximum value of 10.0, on a scale defined by 
the gas disk density. Extreme spatial heterogeneity is evident at
this early time, 112 yr after injection occurs.}
\end{figure}

\clearpage

\begin{figure}
\vspace{-2.0in}
\plotone{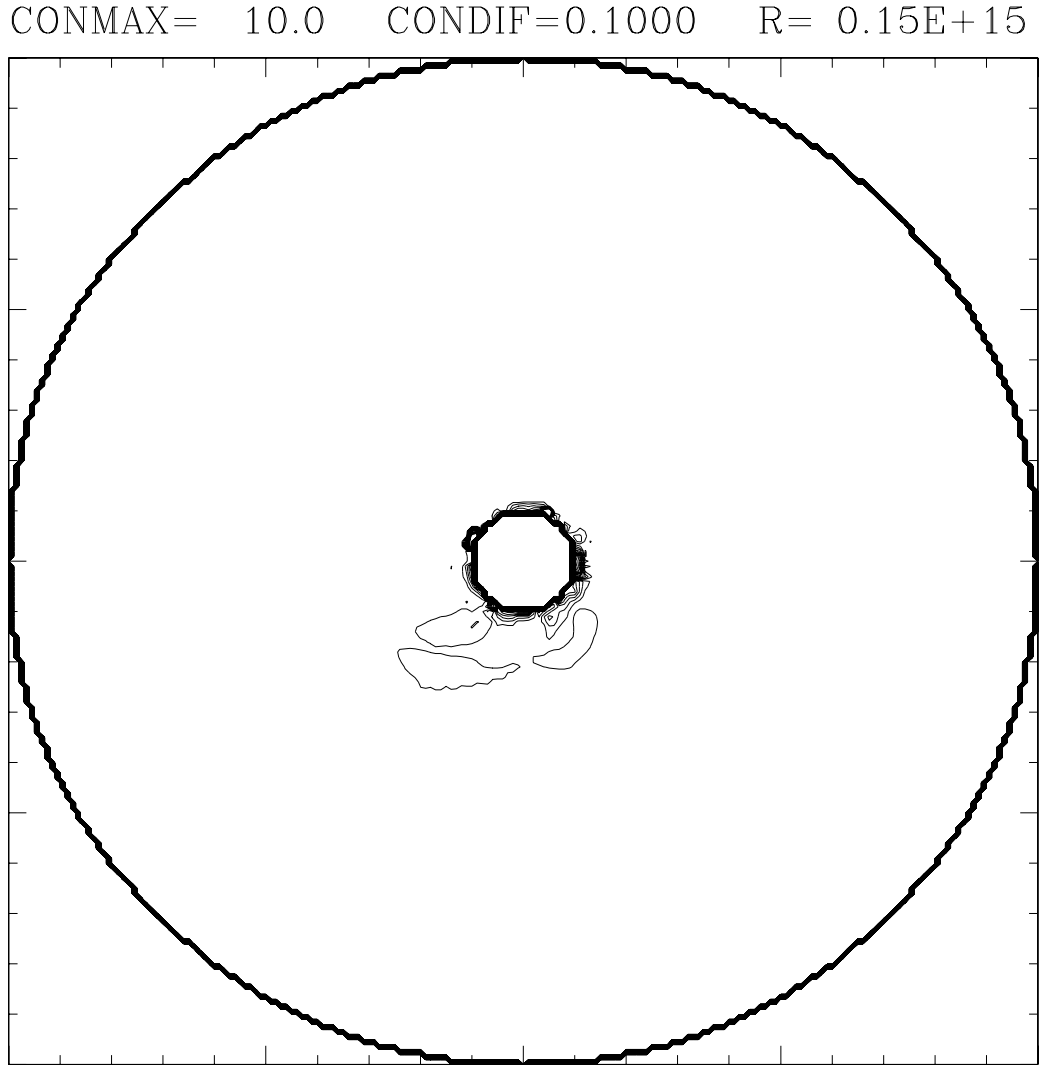}
\caption{Same as Figure 8, except after 1548 yr of evolution of 
model 2.5. The degree of spatial heterogeneity has been reduced
throughout most of the disk to levels less than 25\%.}
\end{figure}

\clearpage

\begin{figure}
\vspace{-2.0in}
\plotone{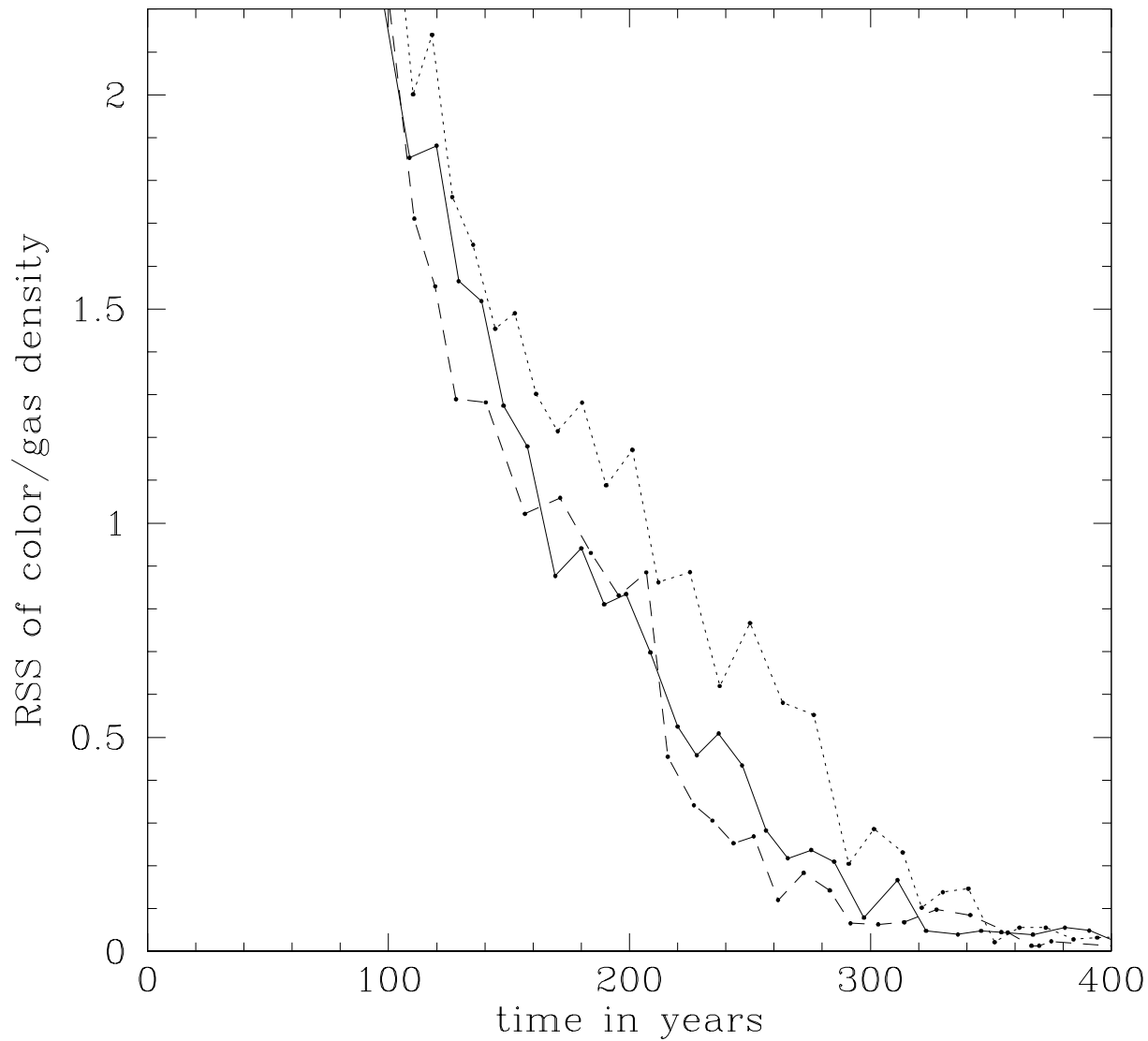}
\caption{Time evolution of the dispersion from the mean (i.e., standard
deviation, or the root of the sum of the squares [RSS] of the differences
from the mean) of the color field density divided by the gas density
(e.g., $^{26}$Al/$^{27}$Al abundance ratio) in the disk midplane in
models 1.4 (solid line), 16 (short-dashed line), and 48 (long-dashed line).
The color field was injected onto the disk surface at a time of 0 yr. 
Starting from high values (RSS $>>$ 1), the dispersion decreases on 
a timescale of $\sim 300$ yrs.}
\end{figure}

\clearpage

\begin{figure}
\vspace{-2.0in}
\plotone{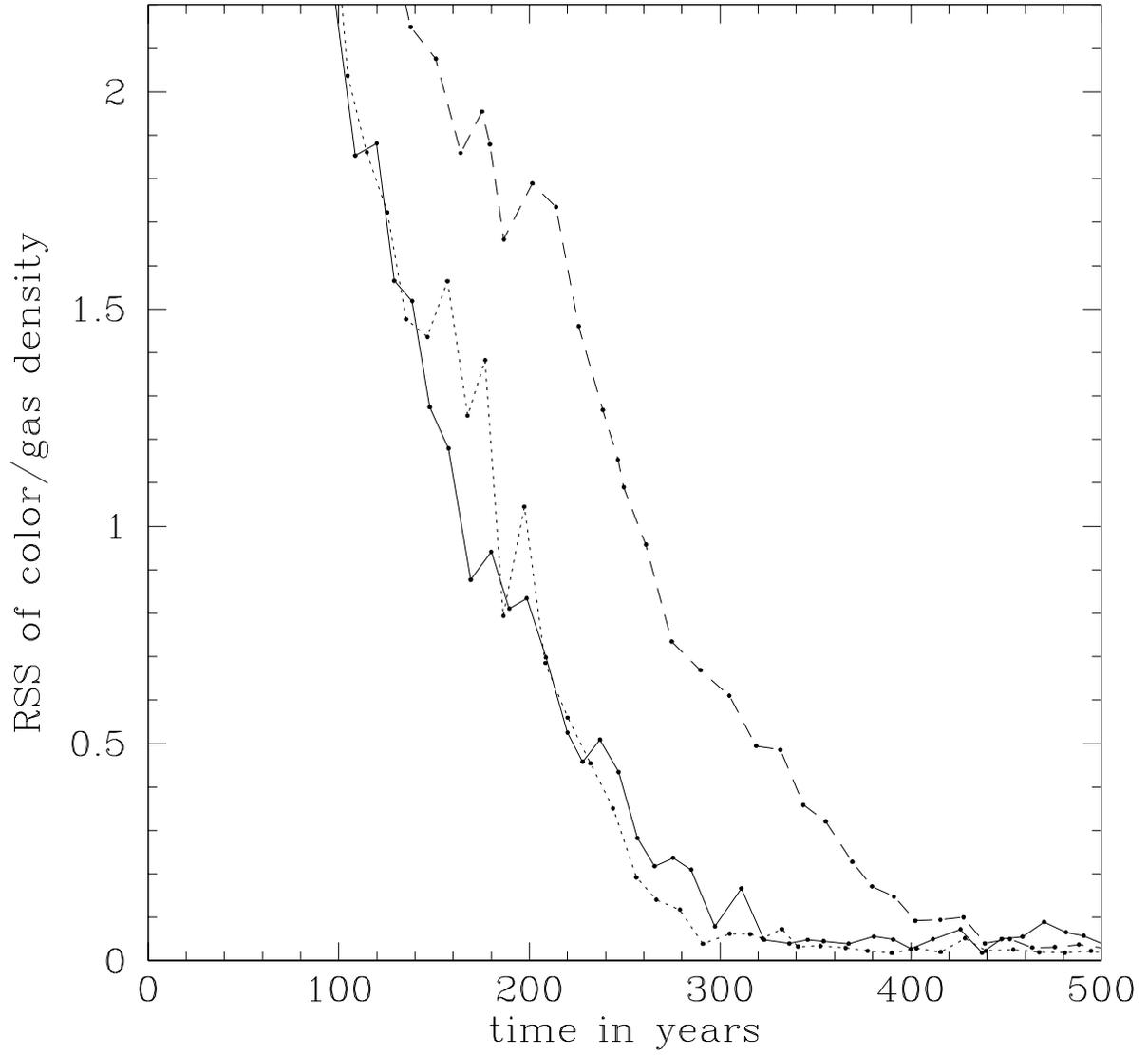}
\caption{Similar to Figure 10, except showing models 1.4 (solid line), 
1.8 (short-dashed line), and 2.5 (long-dashed line).}
\end{figure}

\clearpage

\begin{figure}
\vspace{-2.0in}
\plotone{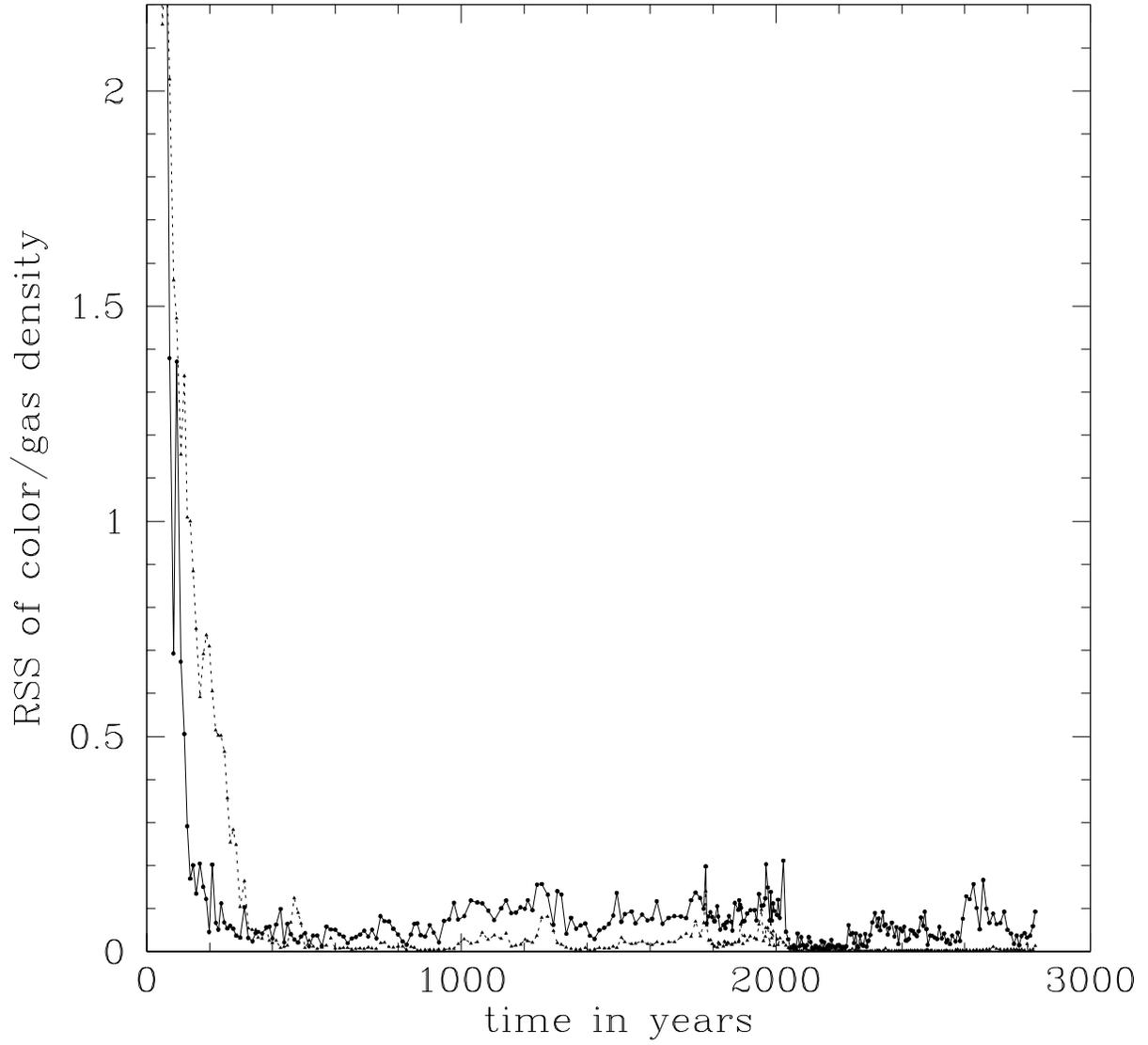}
\caption{Similar to Figure 10, except showing the entire evolution
of model 1.4, with the dispersion displayed separately for the inner
disk (1 to 5.5 AU - solid line) and outer disk (5.5 to 10 AU - dashed
line).}
\end{figure}

\clearpage

\begin{figure}
\vspace{-2.0in}
\plotone{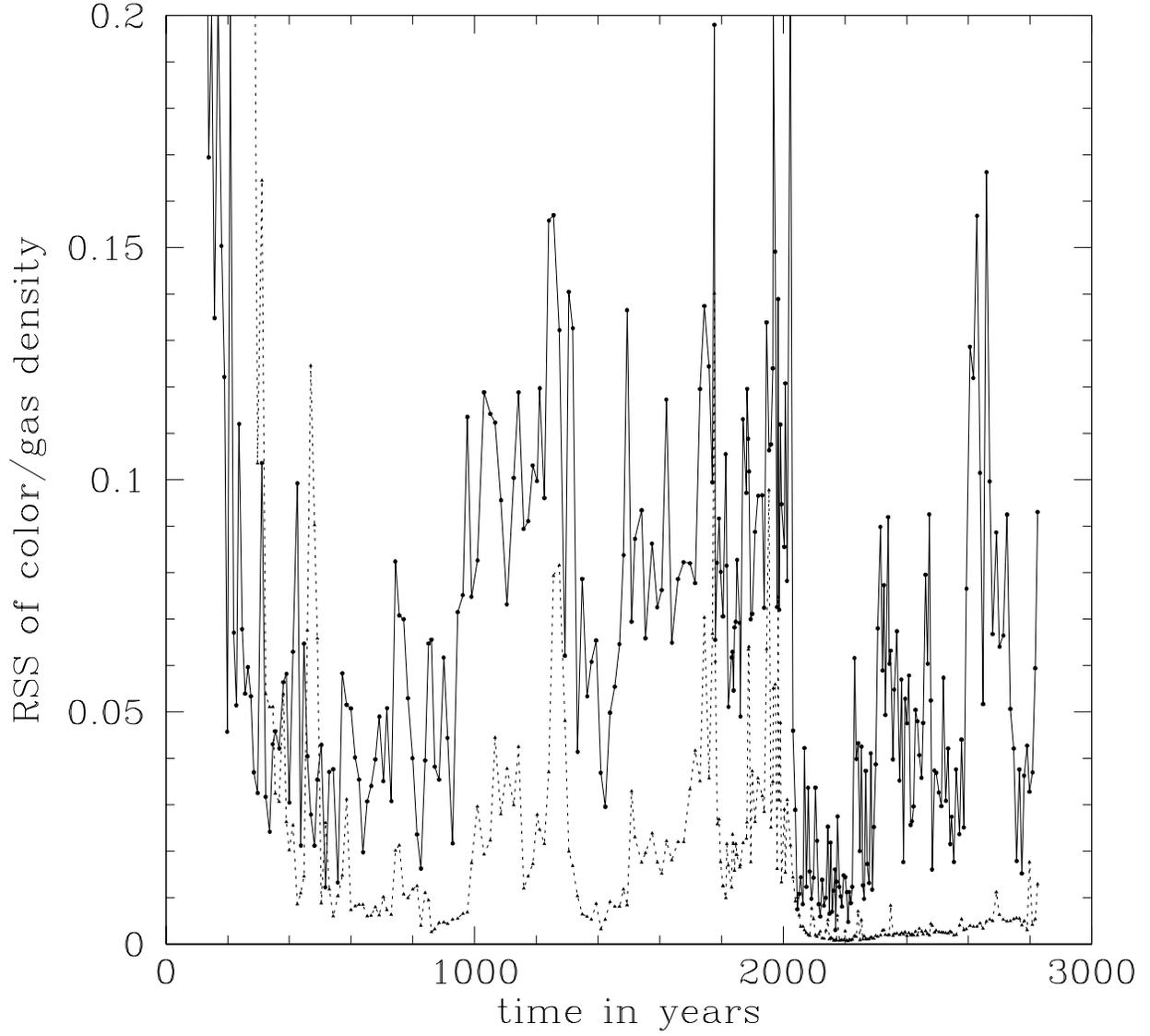}
\caption{Same as Figure 12, but displayed on a different vertical
scale to highlight the later time evolution of model 1.4.}
\end{figure}

\clearpage

\begin{figure}
\vspace{-2.0in}
\plotone{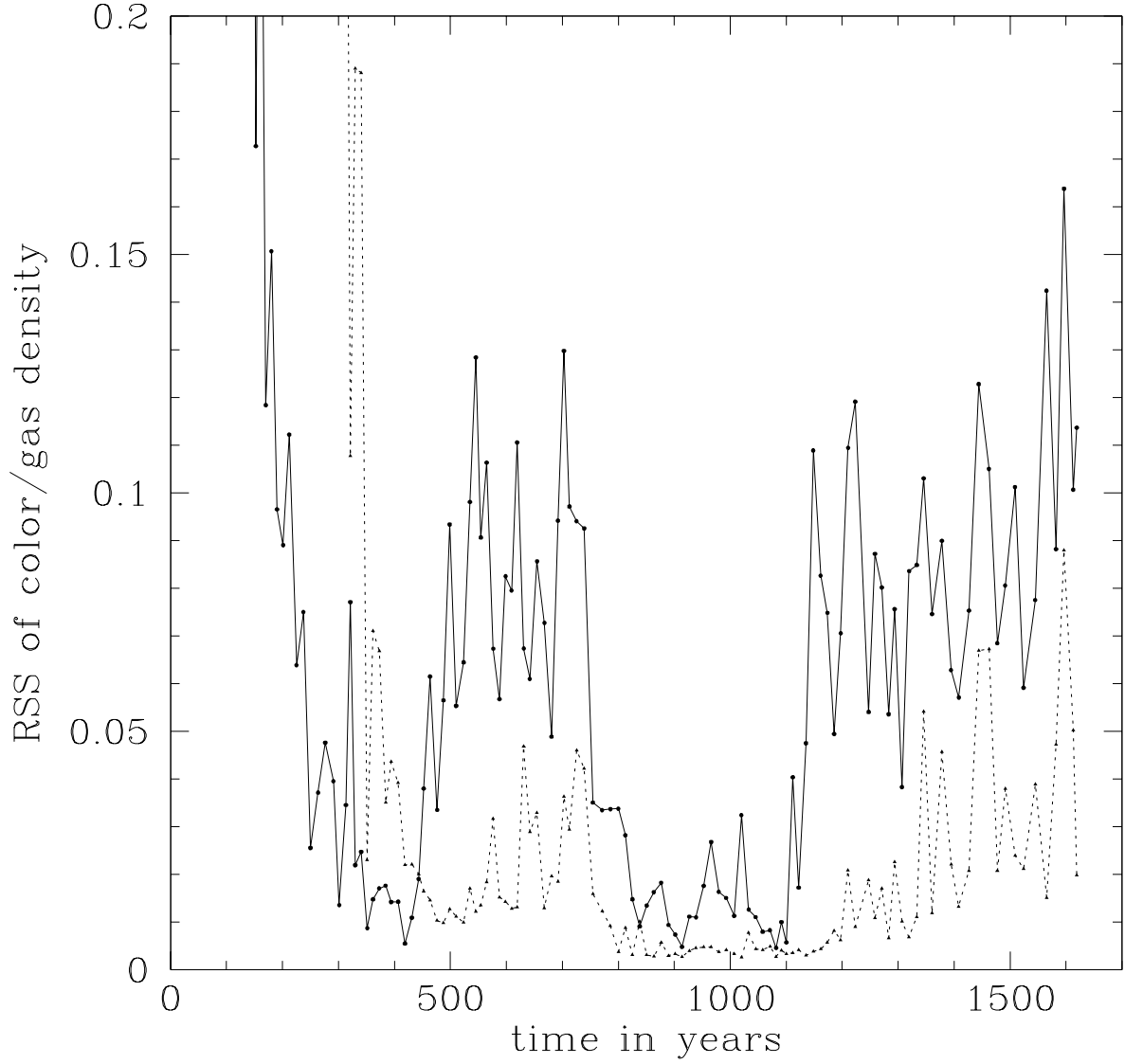}
\caption{Same as Figure 13, but for model 16, showing the evolutions
of the inner disk (solid line) and outer disk (dashed line) 
dispersions.}
\end{figure}

\clearpage

\begin{figure}
\vspace{-2.0in}
\plotone{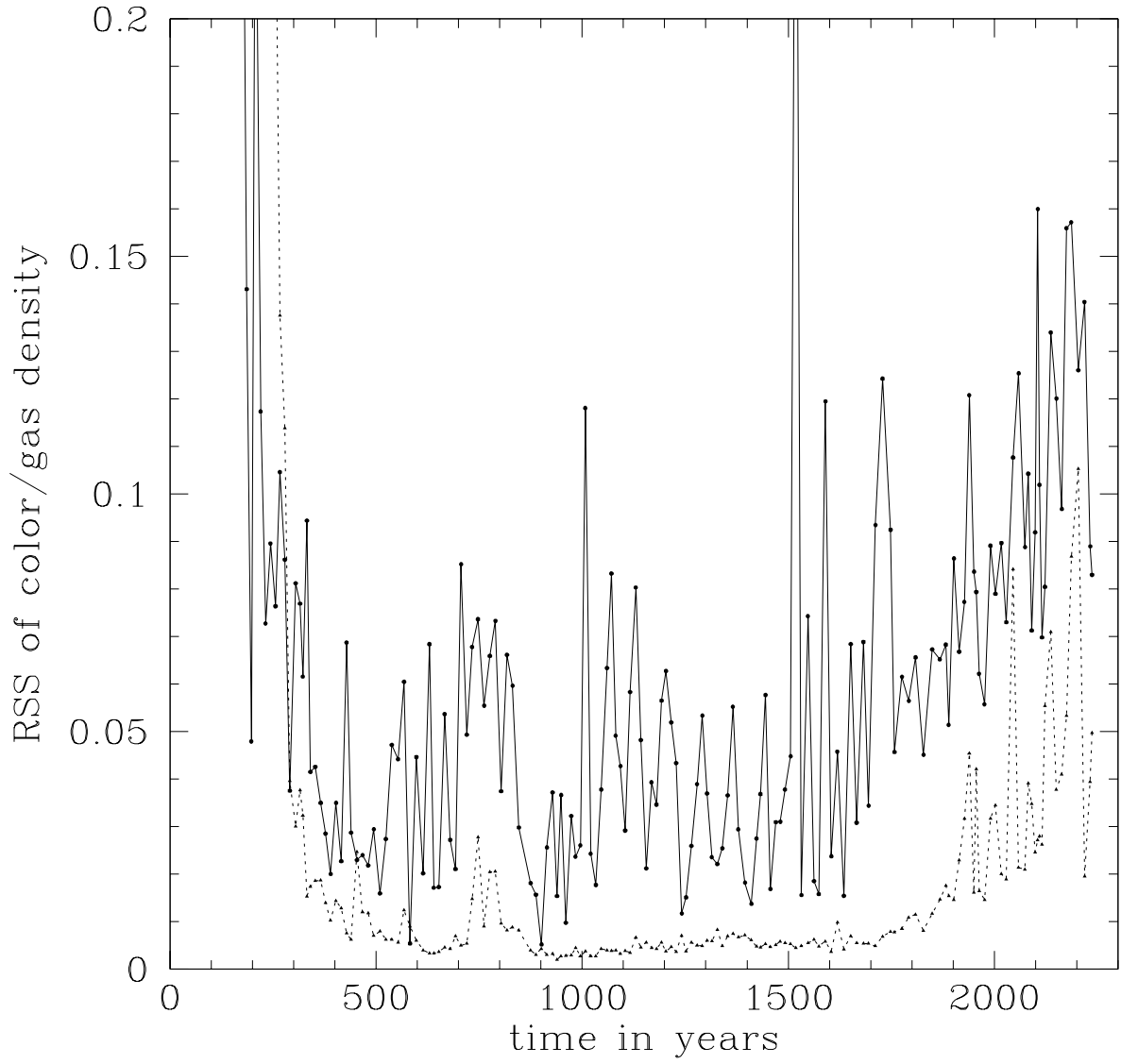}
\caption{Same as Figure 13, but for model 1.8.}
\end{figure}

\clearpage
\begin{figure}
\vspace{-2.0in}
\plotone{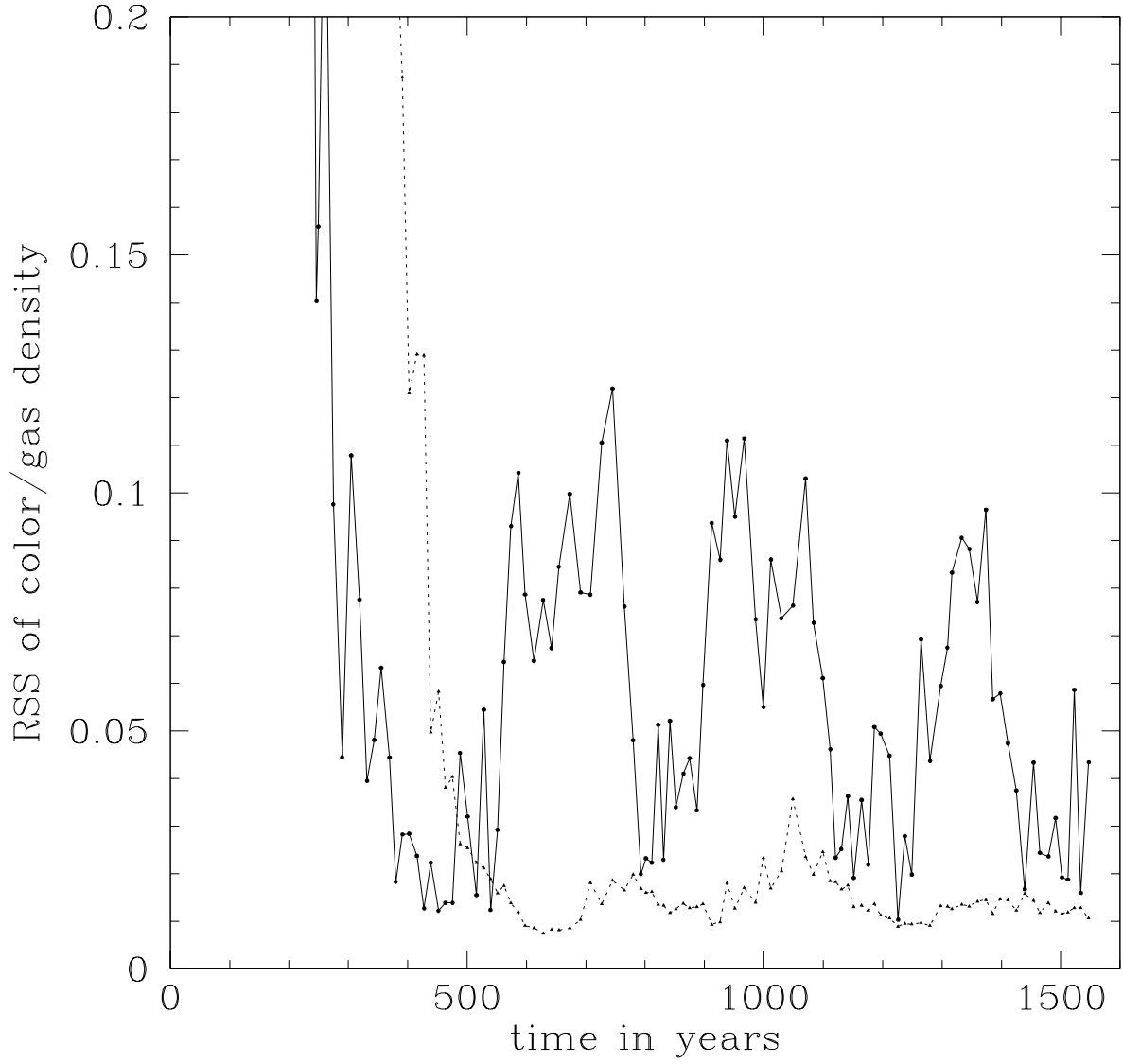}
\caption{Same as Figure 13, but for model 2.5.}
\end{figure}

\clearpage

\begin{figure}
\vspace{-2.0in}
\plotone{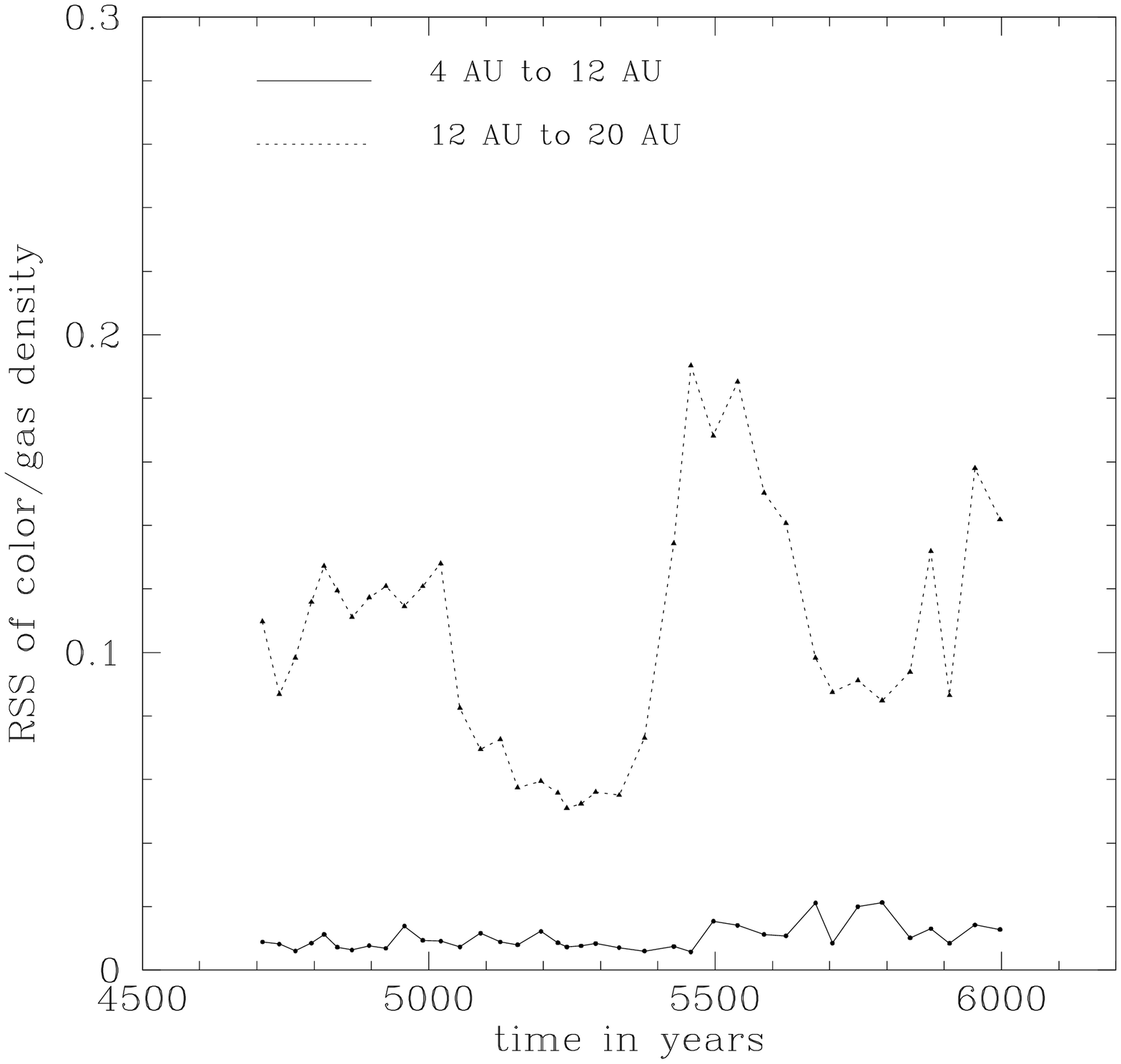}
\caption{Similar to Figure 13, but for model 15S, where the
inner disk now spans from 4 to 12 AU, while the outer disk spans
from 12 to 20 AU.}
\end{figure}

\clearpage

\begin{figure}
\vspace{-2.0in}
\plotone{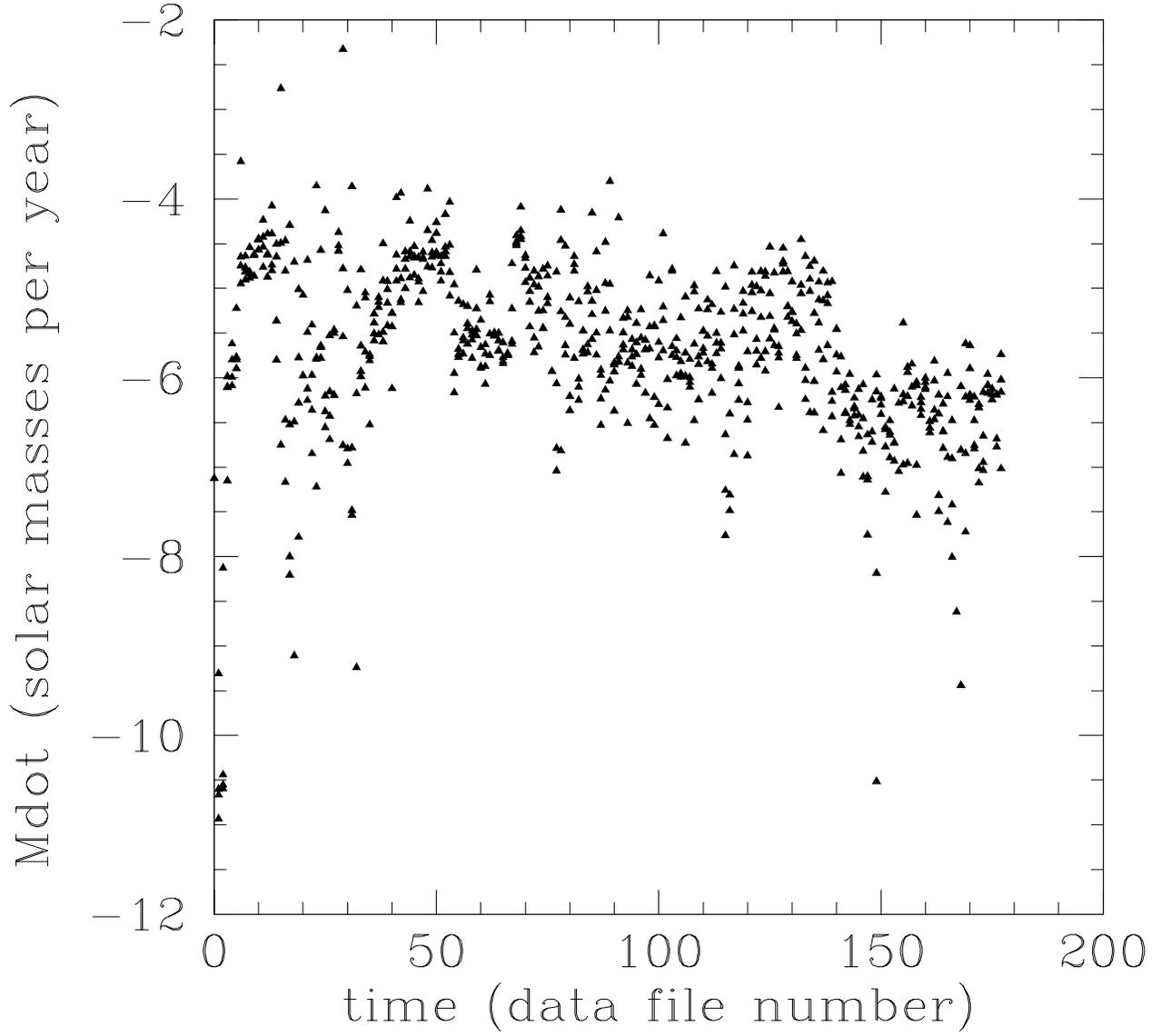}
\caption{Mass accretion rate onto the central protostar as a function
of time for model 1.8. Total time span shown is 2244 yr.}
\end{figure}

\clearpage

\begin{figure}
\vspace{-2.0in}
\plotone{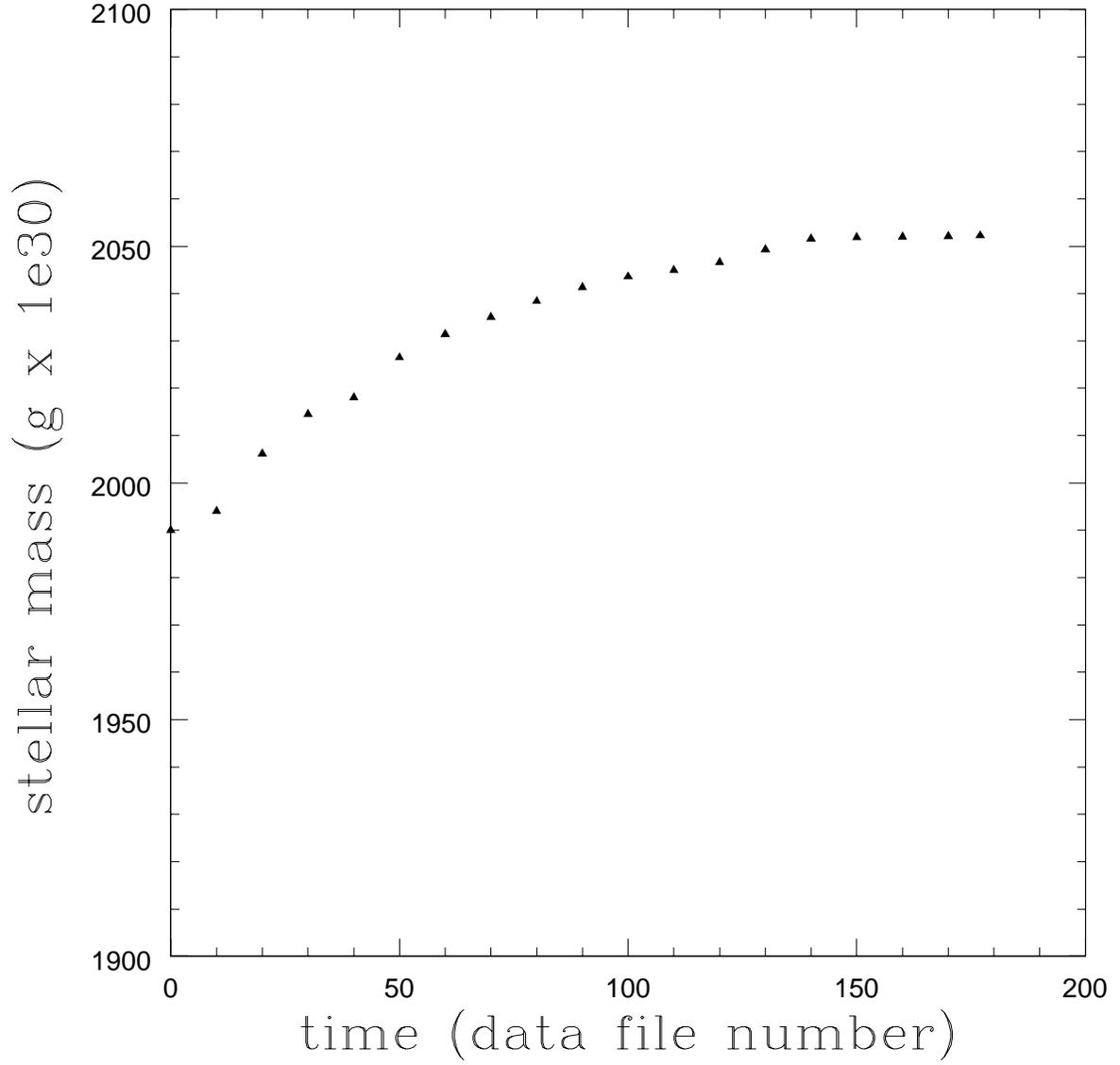}
\caption{Evolution of the mass of the central protostar in model 1.8
as a function of time, spanning 2244 yr, as a result of the mass
accretion rate shown in Figure 18.}
\end{figure}

\clearpage

\end{document}